\documentstyle[12pt]{article}

\topmargin
-2cm
\textwidth
15cm
\textheight
25cm
\oddsidemargin
1cm
\evensidemargin
1cm
\begin{document}
\newcommand{\newc}{\newcommand}
\newc{\sm}{Standard
Model}
\date{}
\title{ String GUTs}
\author{G. Aldazabal \thanks{Permanent Institutions: CNEA, Centro At\'omico
Bariloche, 8400 S.C. de Bariloche, and CONICET, Argentina} $^1$,
A. Font$^2$, L.E. Ib\'a\~nez$^1$ and A. M. Uranga$^1$ \\ \\
$^1$Departamento de F\'{\i}sica Te\'orica, \\
Universidad Aut\'onoma de Madrid, \\
Cantoblanco, 28049 Madrid, Spain.
\\   \\
$^2$Departamento de F\'{\i}sica, Facultad de Ciencias,\\
Universidad Central de Venezuela,\\
A.P. 20513, Caracas 1020-A, Venezuela.\\
}
\maketitle
\vspace{-4in}\hspace{4in} FTUAM-94-28

\hspace{4in}
hep-th 9410206
\vspace{4in}

\begin{abstract}

Standard SUSY-GUTs such as those based on
$SU(5)$ or $SO(10)$ lead to predictions for the values of
$\alpha _s$ and $sin^2\theta _W$ in amazing agreement with
experiment.
In this article we investigate how these models may be obtained
from string theory, thus bringing them into the only known consistent
framework for quantum gravity.
String models with matter in standard GUT representations require
the realization of affine Lie algebras at higher levels.
We start by describing some methods to build level
$k=2$  orbifold string models with gauge groups $SU(5)$ or
$SO(10)$. We present several examples and identify
generic features of the type of models
constructed. Chiral fields  appropriate to break the
symmetry down to the standard model generically appear in
the massless spectrum. However, unlike in
standard SUSY-GUTs, they often behave as string moduli,
i.e., they do not have self-couplings.  We also
discuss briefly the doublet-triplet Higgs splitting.  We find
that, in some models,
built-in sliding-singlet type of couplings exist.

\end{abstract}
\maketitle

\newpage

\section{Introduction}

There has been recently a renewed interest in the
field of supersymmetric grand unified theories
(SUSY-GUTs) \cite{guts}. One of the justifications for that
interest is the very good agreement found between the
predicted \cite{gqw} values for $sin^2\theta _W$ and $\alpha _s$  and the
values experimentally measured, particularly at LEP
\cite{amal}.

We know that SUSY-GUTs by themselves cannot be the whole story
since   they do not address the problem of quantum gravity. On the
other hand, supersymmetric 4-D strings \cite{dine}
provide a general framework
for the unification of all interactions including gravity into
a finite theory.  Thus an  obvious possibility is trying to combine
both elements  and construct 4-D strings whose massless sector
resemble SUSY-GUTs. We will call this type of structure
{\it String GUTs}.

Although this is an obvious idea, in the literature there
are only a few attempts \cite{lew,fiq,nanop,otros,nos} at
developing it. The reason for this situation is twofold:

{\bf i.}  A fundamental ingredient in any SUSY-GUT is the set of chiral
fields that break the GUT symmetry down to the
standard model, the GUT-Higgs fields. In $SU(5)$ the simplest option
is an adjoint 24-plet. In $SO(10)$ one needs two set of Higgses,
apart from the adjoint $45$ (or the symmetric $54$ representation)
one needs other Higgs fields, such as $16+{\bar {16}}$ or $126+{\bar {126}}$,
to lower the rank. However, in the most common type of 4-D strings
built in the past, no adjoints (nor 54s in the $SO(10)$ case)
may be present in the spectrum of the theory since the affine Lie
gauge algebra is realized at level $k=1$. Thus, models with the gauge
group realized at $k\geq 2$ must be considered. This turns out to be
technically non-trivial due to the constraints of modular invariance.

{\bf ii.}
It is not obvious that string-GUTs  give us any
phenomenological improvement
over the $k=1$ 4-D string models already constructed in the past.
Indeed, the fact that the gauge couplings unify at a single scale
is also present in strings \cite{gins}
even without any GUT symmetry present,
i.e. in a $k=1$  4-D string with the SM gauge group.
Another prominent virtue of GUTs, charge quantization, may also be seen as a
consequence of the cancellation of anomalies so that it is not necessary to
invoke any GUT symmetry to explain it.
In addition SUSY-GUTs are not free of some important
problems, in particular the doublet-triplet splitting problem which has
already been with us for more than a decade \cite{guts} .
This problem is on the other hand
not necessarily present in a 4-D string with SM gauge group.

It is perhaps time to reconsider the above two points since the
fantastic agreement between theory and experiment for the joining of
coupling constants may turn out not to be just a coincidence.
 Although couplings are also unified in e.g.
a string with a SM gauge group, the unification scale is
$M_{string}$ \cite{kaplu}
which is around a factor 20 larger than the value
$M_X\simeq 10^{16}$ GeV suggested by the extrapolation of the
low energy coupling constants. This in turn leads to numerical
results for $sin^2\theta _W$ and $\alpha _s$ in disagreement with data
\cite{ilr,anton} .
This factor 20 may be explained
by different effects like large string threshold
corrections \cite{ilr,il} ,
 existence of extra matter fields beyond those of
the MSSM  \cite{anton} or a non-canonical value for the normalization
$k_1$ corresponding  to the $U(1)$ hypercharge
\cite{iki} . Although indeed
all or some of these effects may be present one must admit that then
one is really adjusting, not  predicting, the value of $sin^2\theta
_W$ and $\alpha _s$.

We think that, in view of the above arguments, it is worth attempting to
construct GUTs from strings. Even from the merely technical string point
of view it is interesting to construct higher level 4-D string models
in order to study their generic properties as compared to the
$k=1$ models constructed in the past.
We address in this article the construction of higher level
string GUTs by using 4-D orbifold techniques.
To the best of our knowledge only Ref. \cite{fiq} has previously
dealt with the problem of constructing orbifold models with higher
level affine Lie  algebras. There it is explained how three different
methods can be used to construct 4-D orbifolds with some group factor
realized at higher level. In particular, an specific $Z_3$
  orbifold toy-model with an $SU(3)$ group realized at level
$k=3$  was constructed simultaneously by the same three
methods that will be used in the following.

The structure of this paper is as follows. In section 2 we describe some
model-independent aspects of orbifold string-GUTs that are present
 independently of the construction method used. Some of these may be
easily adapted to other left-right symmetric $(0,2)$ constructions.
We have found that the $Spin(32)/Z_2$ lattice is a better starting
point than $E_8\times E_8$ in order to obtain string-GUTs.
Due to this we include in section 3 a discussion of some general
properties of orbifold models in $Spin(32)/Z_2$.

In the following three sections we construct string-GUTs with the
three methods of \cite{fiq}. In the method of continuous Wilson
lines  used in section 4, the starting point is a $(0,2)$
orbifold in which the embedding in the gauge degrees of freedom
is totally or partially realized through an automorphism of the
gauge lattice (instead of a shift vector). The
projection on invariant states forces the left-moving piece of
some untwisted fields to be combinations  of the usual
$e^{iP\cdot F}$ vertex operators. In particular, Cartan
subalgebra states will involve automorphism invariant
combinations. The next step is the addition of a continuous
Wilson line. Gauge states, including Cartan generators,
are eliminated from the massless spectrum and the rank is reduced,
often leaving behind a subgroup at higher level.

The second method is described in section 5. This involves
the modding by a permutation of identical gauge factors.
The starting point is a level $k=1$ model in which
the  orbifold twist is embedded in the gauge degrees of
freedom through a shift in the gauge lattice. The
orbifold  and the shift are chosen so that the
observable gauge group has repeated factors such as
$SU(5)\times SU(5)$ or  $SO(10)\times SO(10)$.
More generally, gauge groups of the form $G_{GUT}\times {\hat G}$
with $G_{GUT}\subseteq {\hat G}$ are also used.
The
next step is to add a discrete Wilson  line realized as a
permutation of the repeated group factors.  The projection on
invariant states applied to the gauge fields  requires forming
symmetric combinations of the   generators of
the two groups, leading to a surviving diagonal group  realized at
higher level.

In the method of flat directions discussed in section 6, the
starting point is also a level $k=1$ model. In particular
cases where there are scalar field directions flat to all orders,
the original gauge symmetry can be spontaneously broken
to a subgroup which is realized at $k>1$. Unlike the other
two methods that are stringy in nature, this is field-theoretical
since the flat directions are analyzed in terms of the effective
$N=1$ supersymmetric Lagrangian.
A typical base model would contain the gauge group $SU(5)\times SU(5)$
together with massless fields transforming as $(5,{\bar 5})$ and
$({\bar 5},5)$. If there is a flat direction in which these fields
acquire an appropriate Vev, the gauge group is broken to a diagonal
$SU(5)$ realized at level $k=2$. Likewise, $(10,10)$ multiplets
can break $SO(10) \times SO(10)$ to the diagonal $SO(10)$ at
level $k=2$.

In section 7 we discuss different phenomenological aspects of
the class of string-GUTs encountered including a discussion of
the generic features of the $SO(10)$ GUTs found, the structure of
the GUT-Higgs potential, the doublet-triplet splitting problem and
other general features. We present some final comments and an outlook
in section 8. While constructing the specific  string-GUTs we had to
resolve several subtleties concerning the generalized GSO
projectors in orbifolds with Wilson lines as well as other
technical issues which are discussed in the appendix.

\section{General properties of higher level orbifold GUTs}

In this section we discuss general properties of higher level
\cite {km} orbifold constructions. Some of these properties are actually
model-independent and will also apply to other types of 4-D strings.
Our starting point is the ten-dimensional heterotic string
in which the gauge degrees of freedom arise from the extra 16
left-moving coordinates $F_I$ compactified on a torus with
$E_8\times E_8$ or $Spin(32)/Z_2$ lattice. The observable
gauge group $E_8\times E_8$ or $SO(32)$ corresponds to
an affine Lie algebra at level $k=1$ \cite {km} .
A further process of compactification and twisting will generically
lead to a 4-D model with gauge group $ G_1 \times G_2 \times \cdots$
in which each factor is associated with an algebra at level $k_i$.
We recall that for a non-Abelian algebra, $k$ must be a positive
integer whereas for a $U(1)$ factor, $k$ is not really a level
but a real positive normalization constant.
The possible non-Abelian levels are constrained
\cite{lew,fiq,nanop} by the
condition that the total contribution $c_G$ of the gauge sector to the
(left) central charge satisfies $c_G \leq 22$ (or $c_G \leq16$
if there are no enhanced gauge symmetries). More precisely,
\begin{equation}
c_G\ =\ \sum _i c_i =
 \sum _i {{k_idimG_i}\over {k_i+ \rho_i}}\ \leq \ 22
\label{cc}
\end{equation}
where $dim  G_i$ and $\rho_i$ are respectively the dimension
and the dual Coxeter number of $G_i$. In particular,
$\rho=N$ for $SU(N)$, $\rho=2(N-1)$ for $SO(2N)$ and
$\rho=12,18, 30$ for $E_{6,7,8}$. A $U(1)$ factor contributes
1 to the sum.
Condition (\ref{cc}) immediately gives useful information on the
possible levels of interesting GUT groups. For example,
$SO(10)$ or $E_6$ can at most be realized at levels
7 and 4 respectively (4 and 3 if there are no enhanced symmetries).
For simply-laced groups $c_i=rank  G_i$ when $k_i=1$. Since
$c_i$ increases with the level we can also conclude that higher
level models necessarily have lower rank.

A second important constraint concerns the possible particles which
may appear in the string spectrum, both massless and massive.
At level $k$, the allowed unitary highest-weight representations
must satisfy the condition:
\begin{equation}
\sum_{j=1}^{rankG} n_jm_j\ \leq \ k
\label{uni}
\end{equation}
where $n_j$ are the Dynkin labels of the highest weight of the
representation of $G$, and $m_j$ are positive integers ($\leq 6$) known
for every simple lie group. In particular, for $SU(N)$, $m_j=1$,
so that for $k=1$ the only allowed representations are those with Dynkin
levels $(1,0,\cdots )$, i.e., the fundamental and completely antisymmetric
representations. For $SO(2N)$ only the vector and spinor
representations are allowed at $k=1$.
Since adjoint scalars do not appear in the spectrum,
the possibility of constructing GUT-like string models
at level $k=1$ is ruled out.

There are stronger constraints \cite{fiq,nanop}
on the possible particles
which could be present in the $massless$ spectrum. Naively
speaking, in string theory the more quantum numbers a particle has,
the less likely for it to be massless. This is a very important
property of string theories which is often not sufficiently
emphasized. Let us now discuss this property in the
context of orbifold models. The conclusions may be generalized
to other types of 4-D string constructions. The mass formula
for the left-movers of a heterotic 4-D string is given by:
\begin{equation}
{1\over 8}M_L^2\ =\
N_L\ +\ h_{KM}\ +\ E_0\ -1  \ .
\label{ml}
\end{equation}
Here $N_L$ is the left-moving oscillator number,
$h_{KM}$ is the contribution of the gauge sector
to the conformal weight of the particle and $E_0$ is the
contribution of the internal (compactified) sector to the
conformal weight.

Let us consider first the case of symmetric
$(0,2)$ Abelian orbifolds. All Abelian $Z_N$ and $Z_N\times Z_M$
orbifolds may be obtained by toroidal compactifications in
which the 6 (left and right) compactified dimensions are
twisted. There are just 13 possible orbifold twists which can be
characterized by a shift $v=(v_1,v_2,v_3)$, where $e^{2i\pi v_i}$
are the three twist eigenvalues in a complex basis.
The 13 possible shifts are shown in Table \ref{tuno}.
A consistent symmetric orbifold model is obtained by combining
different twisted sectors in a modular invariant way. This
procedure is well explained in the literature \cite{orb2,imnq}.
Let us just mention that a given twist in the table can be present
in several different orbifolds. For example, the shift $(0,1/6,1/6)$
appears in the $Z_{12}$, $Z_3\times Z_6$,
$Z_2\times Z_6$ and $Z_6\times Z_6$ orbifolds.

\begin{table}
\begin{center}
\begin{tabular}{|c|c|c|c|c|}
\hline
$(|v_1|,|v_2|,|v_3|)$
   & $E_0$ & ${\underline {24}}$ & $(5, {\bar 5})$, ${\underline {45}}$
& $(10,10)$,${\underline {54}}$  \\
\hline
$(0,0,0)$  & 0&$\surd$&$\surd$& $\surd $ \\
\hline
$(1/3,1/3,2/3)$ & 1/3 &     &    & \\
\hline
$(1/2,1/4,1/4)$ & 5/16 &  &  & \\
\hline
$(1/3,1/6,1/6)$ & 1/4 &$\surd $ &    & \\
\hline
$(1/2,1/3,1/6)$ & 11/36 &    &    & \\
\hline
$(3/7,2/7,1/7)$ & 2/7  &$\surd $ &   & \\
\hline
$(1/2,1/8,3/8)$ & 19/64  &&  & \\
\hline
$(1/4,1/8,3/8)$ &  17/64 & $\surd $  &    & \\
\hline
$(1/3,1/12,5/12)$ &13/48  & $\surd $  &   & \\
\hline
$(1/2,1/12,5/12)$      & 41/144  & $\surd $  &    & \\
\hline
$(0  ,1/2,1/2)$ & 1/4 &$\surd $ &  & \\
\hline
$(0  ,1/3,1/3)$ & 2/9  &$\surd$&   & \\
\hline
$(0  ,1/4,1/4)$ & 3/16 &$\surd$&$\surd $& \\
\hline
$(0  ,1/6,1/6)$ & 5/36 &$\surd$& $\surd $& \\
\hline
\end{tabular}
\end{center}
\caption{Massless GUT-Higgs fields allowed in the
different twisted sectors of all Abelian orbifolds.}
\label{tuno}
\end{table}

To each possible twisted sector there corresponds
a value for $E_0$ given by the general formula:
\begin{equation}
E_0\ =\ \sum _{i=1}^3\ {1\over 2}|v_i|(1-|v_i|)
\label{ecero}
\end{equation}
Notice also that $E_0=0$ for the untwisted sector
which is always part of any orbifold model.
The value of $E_0$ is also shown in Table \ref{tuno}.
In the case of asymmetric orbifolds, obtaining $N=1$ unbroken
SUSY allows the freedom of twisting the right-movers
while leaving untouched the (compactified) left-movers.
In this case one can then have $E_0=0$ even in twisted sectors.

Let us go now to the other relevant piece in eq. (\ref{ml}),
namely the contribution $h_{KM}$
of the affine algebra sector to the conformal weight of the particle.
If the orbifold twist is just embedded in the gauge degrees of
freedom through a shift $V$, each non-Abelian factor of
the resulting group inherits level $k_i=1$
and furthermore $h_{KM}=(P+V)^2/2$.
More generally, we assume that further action on internal and
gauge degrees of freedom leads to factor groups at higher levels.
A state in a representation $(R_1,R_2, \cdots)$ will then have
\begin{equation}
h_{KM}\ =\ \sum _i {{C(R_i)}\over {k_i+\rho_i}}
\label{peso}
\end{equation}
Here $C(R)$ is the quadratic Casimir of the representation $R$.
$C(R)$ may be computed using $C(R)dim(R)=T(R)dimG$,
where $T(R)$ is the index of $R$. Unless otherwise explicitly stated,
we use the standard normalization in which
$T=1/2$ for the $N$-dimensional representation of $SU(N)$ and
$T=1$ for the vector representation of $SO(2N)$.
With this normalization, for simply-laced groups
the Casimir of the adjoint satisfies $C(A)=\rho.$
The contribution of a $U(1)$ factor to the total
$h_{KM}$ is instead given by $Q^2/k$, where $Q$ is the $U(1)$ charge
of the particle and $k$ is the normalization of the
$U(1)$ generator, abusing a bit it could be called the level
of the $U(1)$ factor.
Formula (\ref{peso}) is very powerful because the
$h_{KM}$ of particles can be computed without
any detailed knowledge of the given 4-D string. This information
is a practical guide in the search for models with
some specific particle content.

In this article we are mainly interested in the construction of
GUT models with gauge groups $SU(5)$ and $SO(10)$ at level $k=2$.
As it will become clear, to this end it is sometimes
useful to look for models of the form $SU(5)\times SU(5)$
and $SO(10)\times SO(10)$ at level  $k=1$. The values of
$h_{KM}$ for the lowest dimensional representations of these
groups are given in Tables \ref{tdosa} and
\ref{tdosb}.
We have also included the equivalent results for
some representations of the $SO(10)$ subgroup $SU(4)\times SU(2)
\times SU(2)$. Notice that the values of $h_{KM}$ given in these
tables should be considered as lower bounds on $h_{KM}$ since in
specific models a given representation, e.g. a $24$ of
$SU(5)$, could be charged with respect to other gauge groups
in the model, e.g. a $U(1)$ factor might be present.

\begin{table}
\begin{center}
\begin{tabular}{|c|c|c|c|c|c|c|}
\hline
${\bf SU(5)}$
   & ${\bf 5}$    &  ${\bf 10} $  &  ${\bf 24 }$ &  ${\bf 15}  $ &
 ${\bf 40}$  & ${\bf 50}$  \\
\hline
${\bf k=1}$ &  2/5  &  3/5   &  -  &  -  &  -  &  - \\
\hline
${\bf k=2  }$
& 12/35        &  18/35   &   5/7  &   28/35   & 33/35  & - \\
\hline
${\bf SO(10)}$
& ${\bf 10}$   &  ${\bf 16} $  &  ${\bf 45 }$ &  ${\bf 54}  $ &
 ${\bf 120}$ & ${\bf 126}$ \\
\hline
${\bf k=1}$ &  1/2  &  5/8   &  -  &  -  &  -  &  - \\
\hline
${\bf k=2}$ &  9/20 &  9/16  &  4/5   &  1      & -   & - \\
\hline
${\bf SU(4)\times SU(2)\times SU(2)}$ &
${\bf (4,1,2)}$ & ${\bf (1,2,2)}$ & ${\bf (6,1,1)}$ & ${\bf (6,2,2)}$ &
 ${\bf (20,1,1)}$ &  ${\bf (1,3,3)}$  \\
\hline
${\bf k=1}$ &  5/8  &  1/2   &   1/2  &  1   &  -  &  - \\
\hline
${\bf k=2}$ &  1/2  &  3/8   &  5/12  & 19/24   &  1      &  1 \\
\hline
\end{tabular}
\end{center}
\caption{ Conformal weights $h_{KM}$ for different representations
of the unifying groups $SU(5)$, $SO(10)$ and $SU(4)\times SU(2)
\times SU(2)$.}
\label{tdosa}
\end{table}

\begin{table}
\begin{center}
\begin{tabular}{|c|c|c|c|}
\hline
${\bf SU(5)\times SU(5)}$
   & ${\bf (5,{\bar 5})}$   & ${\bf (10, 5)}$ & ${\bf (10,10)}$   \\
\hline
${\bf k=1}$ &  4/5  &  1     &   6/5 \\
\hline
${\bf SO(10)\times SO(10)}$
   & ${\bf (10, 10)    }$   & ${\bf (10,16)}$ & ${\bf (16,16)}$  \\
\hline
${\bf k=1}$ &  1    &  9/8   &   10/8 \\
\hline
\end{tabular}
\end{center}
\caption{Conformal weights $h_{KM}$ for different representations
of the unifying groups $SU(5)\times SU(5)$ and $SO(10)\times SO(10)$
($k=1$).}
\label{tdosb}
\end{table}

Using eq. (\ref{ml}), the values for $E_0$ in Table \ref{tuno} and those
for $h_{KM}$ in Tables \ref{tdosa},\ref{tdosb}, we can learn,
for instance, what $SU(5)$ or $SO(10)$
representations may appear in any possible twisted
sector of any given Abelian orbifold. In the case of these
groups we are interested in knowing which twisted sectors
may contain $24$-plets or $45$ and $54$-plets respectively.
In the case of $SU(5)\times SU(5)$ or $SO(10)\times SO(10)$
we need to find out which sectors may contain
$(5,{\bar 5})$'s or $(10,10)$'s respectively. The answer to
these questions is shown
in the last three columns of Table 2 and in Table 3.
For a $24$-plet ($k=2$)
one has $h_{KM}=5/7$; for both $(5,{\bar 5})$ and $SO(10)$
$45$-plets ($k=2$) one has $h_{KM}=4/5$ and, finally,
for both $(10,10)$ and $SO(10)$ $54$-plets ($k=2$) one has
$h_{KM}=1$. From these results we draw the following conclusions:

\bigskip
{\bf i.}
   All representations shown may be present in the untwisted sector
of any orbifold.

\bigskip
{\bf ii.}
  $54$s of $SO(10)$ ($k=2$) and $(10,10)$s of
$SO(10)\times SO(10)$ ($k=1$) can only be present in
the untwisted sector of symmetric orbifolds.

\bigskip
{\bf iii.}
     $(5,{\bar 5})$s of $SU(5)\times SU(5)$ ($k=1$) and
$45$s of $SO(10)$ ($k=2$) may only appear either in the
untwisted sector or else in twisted sectors of the type
$v=1/4(0,1,1)$ or $v=1/6(0,1,1)$. This is a very restrictive
result since Abelian orbifolds containing these
shifts are limited. Notice that the order four shift appears
only in the orbifolds $Z_8,Z_{12}$, $Z_2\times Z_4$ and
$Z_4\times Z_4$. The order six shift is present
in $Z_{12}'$, $Z_2\times Z_6$, $Z_3\times Z_6$ and $Z_6\times Z_6$.

\bigskip
{\bf iv.}
     $24$-plets of $SU(5)$ can never appear in the twisted
sectors of the $Z_3,Z_4,Z_6'$ and $Z_8$ orbifolds.

\bigskip
Table 2 gives us also some extra hints. We observe that
the $54$-plet of $SO(10)$  and the $(10,10)$ of
$SO(10)\times SO(10)$ not only are forced to be in the untwisted
sector but have exactly $h_{KM}=1$. Thus they can potentially
be associated to untwisted moduli (continuous Wilson
lines, in the language of Refs. \cite{inq2,finq}).
This will turn out to be the case in specific orbifold models,
as will be shown in section 4.

{}From the above conclusions it transpires that looking
for models with GUT-Higgs fields in the untwisted sector should be
the simplest option, since they can always appear in
$any$ orbifold. This option has another positive aspect in
that the multiplicity of a given representation in the
untwisted sector is never very large, it is always less or equal
than three in practically all orbifolds and is normally equal
to one in the case of $(0,2)$ models. Proliferation of too many
GUT-Higgs multiplets will then be avoided.

In building models with the GUT-Higgs fields in the untwisted
sector one is naturally led to work with orbifolds on the
$Spin(32)/Z_2$ lattice as we now explain with a simplified argument.
In models based on the $E_8\times E_8$
lattice, the matter fields in the untwisted sector are
either charged with respect to the first $E_8$ $or$
with respect to the second but there are $no$ untwisted matter
fields which may be charged with respect to both. Thus, if one has
a GUT group of the form $G\times G$, there will not be untwisted
matter fields transforming as $(R,R)$
because each $G$ factor necessarily lies in a different $E_8$.
But this type of matter is in general needed,
at least in the first and third methods above, in order
to obtain a diagonal GUT with gauge group $G$ at $k=2$.

Surprisingly enough, although 4-D orbifolds on the $E_8\times E_8$
lattice have been exhaustively classified and analysed,
we are not aware of any general analysis
of 4-D orbifold strings based on the $Spin(32)/Z_2$  lattice.
Since these compactifications have inherently interesting
properties we will briefly discuss them in the following section.

\section{Abelian orbifolds on the  $Spin(32)/Z_2$ lattice}

Of course, orbifolds are constructed in the same way both
on $E_8\times E_8$ and on $Spin(32)/Z_2$, the only difference
being that in the latter case the gauge lattice consists of
16-dimensional vectors of the form
\begin{eqnarray}
& (n_1,n_2, \cdots ,n_{16}) & \nonumber \\
& (n_1+\frac{1}{2}, n_2+\frac{1}{2},\cdots ,n_{16}+\frac{1}{2}) &
\label{lat}
\end{eqnarray}
with integers $n_i$ satisfying $\sum n_i=even$.
Since the lattice, denoted $\Lambda_{16}$, is self-dual,
any $P \in \Lambda_{16}$ has $P^2=even$. One important
practical difference with the $E_8\times E_8$ case is that
the shorter spinorial weights have $P^2=4$. Hence, $SO(2N)$
spinorial representations cannot appear in the untwisted sector.
Likewise, exceptional observable groups like $E_6$ and $E_7$
are not possible.

The modular invariance constraints on the possible gauge embeddings are
the usual ones. If we associate to a $Z_N$ twist $v$ a
corresponding shift $V$ in $\Lambda_{16}$,
modular invariance of the partition function dictates:
\begin{equation}
N\ (V^2\ -\ v^2)\ = \ 0\ mod \ 2   \ .
\label{Vcond}
\end{equation}
Also, $NV \in \Lambda_{16}$.
In the case of a $Z_M\times Z_N$ orbifold \cite{znzm}
($M \leq N$) with twists $a,b$
realized through gauge shifts $A,B$ one has:
\begin{eqnarray}
M \  ( A^2 - a^2)  & = & 0\ mod\ 2  \nonumber \\
N \  ( B^2 - b^2)  & = & 0\ mod\ 2  \nonumber \\
M \  ( A\cdot B - a\cdot b)  & = & 0\ mod\ 2
\label{ABcond}
\end{eqnarray}
Also, $MA, NB \in \Lambda_{16}$.
In the presence of a discrete Wilson line $L$, the effective lattice
shift becomes  $V+nL$  with $n$ depending on the
particular element of the space group considered
(see Refs. \cite{imnq,inq2,fiqs}
and the Appendix for more details).
The embedding of the twist and discrete Wilson
lines in the gauge degrees of freedom may also be realized by
automorphisms of $\Lambda_{16}$ instead of shifts. In this
case modular invariance restricts the possible automorphisms
allowed as will be exemplified in the next section.

In any orbifold there is always a modular invariant lattice shift
\cite{orb2}
that corresponds to embedding the orbifold shift into an $SO(6)\ \in
SO(32)$. This standard embedding gives $(2,2)$ models and
basically amounts to setting
$V=v$ in $Z_N$ and $A=a$, $B=b$ in $Z_M \times Z_N$.
Whereas in $E_8\times E_8$ the standard embedding leads to
$E_6$ theories which are chiral, in the $Spin(32)/Z_2$ case
the resulting theories have the non-chiral uninteresting gauge group
$SO(26)$. This is the sole reason why
compactifications on $\Lambda_{16}$ have been essentially
ignored in the literature. While this is a sensible
attitude towards $(2,2)$ compactifications,
more general $(0,2)$ theories,
that are in fact normally the case, deserve more attention.
Appropriate embeddings on $\Lambda_{16}$ do lead to $(0,2)$
models that are more suitable in our approach to constructing
standard GUTs.

As we said, in the $E_8\times E_8$ case we have the practical
knowledge of an embedding, the standard one, that is always modular
invariant for {\it any } orbifold and leads to a chiral model.
We do not know of an embedding in $Spin(32)/Z_2$ which is
modular invariant {\it for any orbifold} and leads to a chiral
model. However we have found that both for $Z_N\times Z_M$
and $Z_N$ orbifolds on $\Lambda_{16}$ there is a
natural embedding which we call the {\it five-fold standard embedding}
that leads to a chiral model and is {\it almost always}
modular invariant. Furthermore, it naturally provides for
$SU(5)$ and $SO(10)$ unification in the same sense that
the usual standard embedding provides for $E_6$ unification
in the $E_8\times E_8$ case.

Let us now explain the idea behind this
 five-fold standard embedding.
Consider the $SO(6)$ tangent group of the compactified space
and its subgroup $SO(2)\times SO(2)\times SO(2)$. We want to
embed the latter in a symmetric way into $SO(32)$. This motivates us
to consider the subgroup $SO(10)\times SO(10)\times SO(10)\times U(1)_A$
of $SO(32)$ and associate:
\begin{equation}
SO(2)\times SO(2)\times SO(2)\ \in SO(6)
\longrightarrow \ SO(10)\times SO(10)\times SO(10)\ \in SO(32)
\label{5fold}
\end{equation}
To implement the embedding we associate to an
$SO(6)$ shift $v$ an SO(32) shift $V$ as follows:
\begin{equation}
v\ =\ {1\over N}(a,b,c)\ \longrightarrow  \ V\ =\ {1\over N}
      (a,a,a,a,a,b,b,b,b,b,c,c,c,c,c,d)
\label{5exp}
\end{equation}
It is now obvious why we call it five-fold standard embedding,
it contains five times the standard embedding shift.
The last integer $d$ is associated to the remaining $U(1)_A$
symmetry and its value is fixed by modular invariance. Also,
$NV$ must belong to $\Lambda_{16}$.

We now consider explicitly the case of $Z_M\times Z_N$ orbifolds.
We recall that the possible values of $M$ are $M=2,3,4,6$ and
$N=\alpha M$ for some $\alpha=1,2,3$.
The five-fold embedding of the shift $a=\frac{1}{M} (1,0,-1)$ is
then given by
\begin{equation}
A=\frac{1}{M} (1,1,1,1,1,0,0,0,0,0,-1,-1,-1,-1,-1,d_a)
\label{avec}
\end{equation}
A suitable value of $d_a$ can always ensure the corresponding
modular invariance constraint for all $M$. A twist of this type
leaves a $N=2$ unbroken SUSY. The shift $A$ in (\ref{avec})
implies a gauge group $SU(10)\times SO(10)\times U(1)^2$,
enhanced to $SO(20)\times SO(10)\times U(1)^2$ in the $Z_2$
case. To further reduce to $N=1$ supersymmetry one considers
the simultaneous shift $b=\frac{1}{N}(0,1,-1)$. Its five-fold
embedding is given by
\begin{equation}
B=\frac{1}{N} (0,0,0,0,0,1,1,1,1,1,-1,-1,-1,-1,-1,d_b)
\label{bvec}
\end{equation}
It is easy to check that the above shifts $A$ and $B$ satisfy
the conditions in (\ref{ABcond}) for all $M$ and $N$ by choosing,
for example, $d_a=4$ and $d_b=8$.

The generic gauge group of the $Z_M\times Z_N$ five-fold embedding is
\begin{equation}
SU(5)\times SU(5)\times SU(5)\times U(1)^3\times U(1)_A
\label{5gr}
\end{equation}
It is enhanced to $SO(10)^3\times U(1)_A$ in $Z_2\times Z_2$ and to
$SU(5)^2\times SO(10)\times U(1)^2\times U(1)_A$ in $Z_2\times Z_4$.
The Abelian factor $U(1)_A$ is anomalous and in all the cases
studied its anomaly is cancelled in the usual way by the
4-D version of the Green-Schwarz mechanism. This will turn out
to have important consequences for the one-loop stability of
the string vacua that we will be considering.

The untwisted matter content has also some interesting features
common to all resulting models.
In particular, it contains the multiplets
\begin{eqnarray}
[ (5,{\bar 5},1)\ +\ ({\bar 5},1,5)\ +\ (1,5,{\bar 5}) ] \nonumber \\
+\ 2 [ (5,1,1)\ +\ 2(1,5,1)\ +\ 2(1,1,5) ]
\label{umat}
\end{eqnarray}
where, depending on the particular model, there may be
flippings $5\leftrightarrow {\bar 5}$ in some $SU(5)$
factor. In some cases there may also appear {\it additional}
untwisted matter fields such as 10-plets in $Z_3\times Z_3$.
The interesting point is that fields with these characteristics
will be important in obtaining appropriate GUT-Higgs fields
as we will see later on.

In $Z_N$ orbifolds the same construction essentially applies. The
five-fold embedding of $Z_3,Z_4,Z_6,Z_7$ and $Z_{12}$ orbifolds
is modular invariant with $d_v=0$, whereas $d_v=4$ is
necessary for $Z_6'$ and $Z_{12}'$. For $Z_N$ orbifolds of small $N$
 one single shift is not enough to achieve all
the  breaking down to $SU(5)^3\times U(1)^3$. For
example, this group is   enhanced to $SU(15)\times U(1)^2$ in
$Z_3$.  Further addition of Wilson lines
would in general be needed to arrive at the smaller
group.

There is an exception to the universal validity
of the five-fold embedding, for the orbifold of order 8
there is no possible choice of $d_v$ that renders this
embedding modular invariant. In spite of this lack of generality,
the five-fold embedding is interesting since it leads in a
natural way to gauge groups which are of phenomenological
interest. Furthermore, a natural replication of these groups occurs.

We wish to emphasize that there are many other embeddings
 in $\Lambda_{16}$ that lead to GUT groups, whether
repeated   or not. We will exploit several possibilities. In
particular,  we will also consider embeddings through
automorphisms of the  gauge
lattice.

Unlike automorphisms of the $E_8$ lattice \cite{e8},
those of $\Lambda_{16}$
have not been studied in any detail. We now give a simplified
analysis adapted to our future needs. We are mostly interested
in automorphisms of order 2 and 4. It is easy to see that two
simultaneous sign flips $F_I \to -F_I$  or two simultaneous
$\pi/2$ rotations $F_I \to F_J$ , $F_J \to -F_I$, are allowed
automorphisms of $\Lambda_{16}$. We use these transformations
as basic building blocks.

Modular invariance, or equivalently left-right level-matching
further restricts the allowed automorphisms. More precisely,
for a $Z_N$ automorphism $\Theta$ we must have
\begin{equation}
E_{\Theta} + E_0 - 1 = 0 \ mod \ \frac{1}{N}
\label{levmat}
\end{equation}
where $E_{\Theta }$ is the vacuum energy shift due to
the $\Theta $-rotated $F$-coordinates.
Notice that $E_{\Theta}$ can be computed by a formula similar
to (\ref{ecero}). For instance, we find $E_{\Theta}=p/8$
for a $Z_2$ automorphism in which $2p$ coordinates change sign.
Such gauge action can accompany an internal shift
$v=(\frac{1}{2},0, -\frac{1}{2})$ with $E_0=1/4$ provided $p=2,6$.
Likewise, for a $Z_4$ automorphism in which $2r$ coordinates
change sign and $4s$ coordinates are rotated by $\pi/2$, we
find $E_{\Theta}=(3s+2r)/16$. This automorphism can then
act as embedding of the shift $v=(\frac{1}{4}, \frac{1}{4},
-\frac{1}{2})$ with $E_0=5/16$ provided $s=1$, $r=0,2,4,6$
or $s=3$, $r=1$.

Lattice shifts equivalent to a given automorphism can also be
determined. For example,
\begin{equation}
 \Theta (F_1,F_2,\cdots ,F_{16}) =
(-F_1,-F_2,-F_3,-F_4,F_5,\cdots , F_{16})  \nonumber
\end{equation}
is equivalent to
\begin{equation}
V= (\frac{1}{2}, \frac{1}{2}, 0, \cdots, 0) \nonumber
\end{equation}
Similarly,
\begin{equation}
 \Theta (F_1,F_2,\cdots ,F_{16}) =
(F_2,-F_1,F_4,-F_3,F_5,\cdots , F_{16})  \nonumber
\end{equation}
is equivalent to
\begin{equation}
V= (\frac{1}{4}, \frac{1}{4}, \frac{1}{2}, 0, \cdots, 0) \nonumber
\end{equation}
The above results can be verified by comparing the spectrum in the
two formulations.

\section{GUTs from continuous Wilson lines: the GUT-Higgs as a
string modulus}

The method of continuous Wilson lines was first introduced in
Refs. \cite{inq2,imnq} as a stringy procedure to reduce the
rank of the gauge group in 4-D orbifold models. Its relationship
with the stringy Higgs mechanism was analyzed in
Refs. \cite{finq,illt}
and recently \cite{wlmod}
a classification of the untwisted moduli space in the case of
$E_8\times E_8$ was worked out for the models obtained using this
method. In Ref. \cite{fiq}
it was explicitly shown how under some circumstances
it also leads to higher level orbifold models. Below we review
its basic features in the case of $Z_N$ orbifolds.

The method relies on the non-Abelian embedding of the orbifold
space group with elements $(\theta, n_ie_i)$, where $\theta$
is the orbifold twist and $e_i$ is an
internal six-dimensional lattice vector. The associated action
is given by $(\Theta,n_i L_i)$, where $\Theta$ is an order $N$
automorphism and $L_i$ is a translation of the gauge lattice.
In the absence of Wilson lines $L_i$, the action
of $\Theta$ can be described by an equivalent shift $V$.
In the presence of $L_i$, the embedding is non-Abelian when
$\Theta L_i$ does not give back $L_i$ up to lattice vectors.
In the following we consider the case of $L_i$ completely
rotated by $\Theta$ so that the condition $(\Theta,n_iL_i)^N=(1,0)$
is automatically fulfilled. This implies that the Wilson lines
$L_i$ are not quantized but may take arbitrary real values and
be continuously varied.

When embedding by automorphisms, not all Cartan gauge currents
are given by combinations of derivatives $\partial F_I$ since
the lattice coordinates $F_I$ are generically rotated by $\Theta$
and the unbroken gauge currents must be invariant under $\Theta$.
The Cartan sub-algebra, as well as the step currents, now arise
from $\Theta$ invariant orbits of the $e^{iP\cdot F}$
operators of the form
\begin{equation}
|P\rangle\ +\ |\Theta P\rangle\ +\ \cdots \ +|\Theta ^{N-1}P\rangle
\label{invs}
\end{equation}
where $|P\rangle \equiv e^{iP\cdot F}$ and $P^2=2$.
On the other hand, untwisted matter states will involve
combinations of the form
\begin{equation}
|P\rangle\ +\ \delta |\Theta P\rangle\ +\ \cdots \ +\
\delta^{N-1}|\Theta ^{N-1}P\rangle
\label{unts}
\end{equation}
that acquire a phase $\delta$, $\delta^N=1$, under $\Theta$.
This phase compensates for the transformation of the right-moving
piece of the full vertex. Combinations of $\partial F_I$ states
can also give rise to untwisted matter.

After the continuous Wilson lines are turned on, states not
satisfying $P.L_i=int$ drop out from the spectrum. This
projection kills some Cartan generators thus forcing a reduction
of the rank of the gauge group. This is a necessary condition
to get a residual algebra realized at higher level.

Concerning the twisted sectors of the orbifold, the left-handed
mass formula now becomes
\begin{equation}
\frac{1}{8} M_L^2 = {1\over 2}(P_T+n_i{L_i}_T)^2\ +\ N_L\
+\ E_0\ +\ E_{\Theta } \ - \ 1
\label{ttwi}
\end{equation}
where $E_{\Theta }$ is the vacuum energy shift due to $\Theta $.
$P_T$ and ${L_i}_T$ are the components of $P$ and $L_i$ which are
left unrotated by $\Theta $. Notice that there are is no winding
in the rotated directions. Also, $N_L$ can now take fractional
values both due to the rotated
$F$-coordinates and the compactified dimensions. States in
twisted sectors organize into representations whose
dimensionality depends on the degeneracy factor
\begin{equation}
D = \sqrt{ \frac  {\det'(1-\Theta )}{|I^*/I|} }
\label{deg}
\end{equation}
where $\det'$ is evaluated in the rotated piece of the lattice.
$I$ is the sub-lattice left invariant by $\Theta $, and $|I^*/I|$ is the
index of its dual $I^*$ on $I$. This factor is similar to that
appearing in asymmetric orbifolds \cite{nsv}
because the gauge twisting
$\Theta $ is asymmetric in nature. Since $\Theta$ does not affect
any right-movers, $D$ is roughly speaking the square root
of the number of points fixed under $\Theta $.
These fixed points in general belong to the unbroken
group weight-lattice and therefore are non-trivially charged.
This means that the degeneracy factor corresponds to some
(reducible) representation of the unbroken gauge group \cite{nsv}.

Before building explicit models let us comment that
since the gauge piece is a complicated linear combination,
it is often difficult to quickly identify the representation and
quantum numbers of a given massless state.
To this purpose it proves convenient to use a parallel
description of the original orbifold, without Wilson lines $L_i$,
in terms of the shift $V$ equivalent to the action of $\Theta$.
In this way, gauge quantum numbers can be more easily determined.

To illustrate the continuous Wilson line method we are going to
build an $SO(10)$ GUT realized at level $k=2$. We will
consider the simplest symmetric orbifold with order 2 symmetries,
namely, $Z_2\times Z_2$.
The internal six-dimensional twists $\theta$ and $\omega$
are respectively realized by
the order two automorphisms $\Theta $ and $ \Omega $ defined by :
\begin{eqnarray}
&  \Theta (F_1,F_2,\cdots ,F_{16})\ =\
(-F_1,-F_2,\cdots ,-F_{10},-F_{11},-F_{12},F_{13},F_{14},F_{15},F_{16}) &
\nonumber \\
& \Omega (F_1,F_2,\cdots ,F_{16})\ =\
(-F_1,-F_2,\cdots ,-F_{10},F_{11},F_{12},-F_{13},-F_{14},F_{15},F_{16})&
\label{auti}
\end{eqnarray}
Notice that $\Theta\Omega$ is another $Z_2$ allowed automorphism.
The unbroken gauge currents correspond to states $|P\rangle$ with
$P$ invariant plus the oscillators $\partial F_{15},\partial F_{16}$.
Also, from non-invariant $P$'s we can form orbits invariant
under both $\Theta$ and $\Omega$. Altogether we find 200 currents
that can be organized into an $SO(10)\times SO(18)\times U(1)^2$
 algebra realized at level $k=1$.

The untwisted matter includes states transforming as
$(10,1)$, $(1,18)$, $(10,18)$ and singlets $(1,1)$. In
the twisted sectors we find matter in the representations
$(16,1)$, $(\overline {16},1)$, $(10,1)$, $(1,18)$ and $(1,1)$,
in multiplicities according to the projectors in (\ref{abcpro}).
At this stage, the model can be equivalently derived
through the shifts
\begin{eqnarray}
A & = & ({1\over 2},{1\over 2},{1\over 2},{1\over 2},{1\over 2},
{1\over 2},0,0,0,0,0,0,0,0,0,0)  \nonumber\\
B & = & ({1\over 2},{1\over 2},{1\over 2},{1\over 2},{1\over 2},
0, {1\over 2},0,0,0,0,0,0,0,0,0)
\end{eqnarray}
In this formulation all charges can be easily determined.

Next we turn on a Wilson line background $L$ along, say,
the compactified direction $e_6$. $L$ has the form
\begin{equation}
L\ = \ (\lambda , \lambda  ,\lambda , \cdots ,\lambda ,0,0,0,0,0,0)
\label{wil}
\end{equation}
The parameter $\lambda $ can take any real value
since $L$ is completely rotated by both $\Theta $ and $\Omega $.
The gauge group is broken to $SO(10)\times SO(8)\times U(1)^2$. The
associated currents are
\begin{eqnarray}
 &{\bf SO(10)} & \nonumber\\
	& |{\underline{+1,-1,0,0,\cdots ,0}},0,0,0,0,0,0\rangle\ +\
|{\underline{-1,+1,0,0,\cdots ,0}},0,0,0,0,0,0\rangle &
\nonumber\\
&{\bf SO(8)}  &  \nonumber\\
& |0,0,0,0,0,0,0, 0 ,0,0,0,0,0,0,\pm 1 ,\pm 1\rangle & \nonumber\\
&\partial F_{15} \  ,   \ \partial F_{16} & \nonumber\\
&|0,\cdots ,0,0,0,{\underline {+1,0}},{\underline {\pm 1,0}}\rangle\ +\
|0,\cdots ,0,0,0,{\underline {-1,0}},{\underline {\pm 1,0}}\rangle &
\nonumber\\
& |0,\cdots ,0,{\underline {+1,0}},0,0,{\underline  {\pm 1,0}}\rangle\ +\
|0,\cdots ,0,{\underline {-1,0}},0,0,{\underline  {\pm 1,0}}\rangle&
\nonumber\\
&\big [ |0,\cdots ,0,{\underline {+1,0}},{\underline {+1,0}},0,0\rangle\  + \
|0,\cdots ,0,{\underline {-1,0}},{\underline {+1,0}},0,0\rangle\ + &\nonumber\\
& \ |0,\cdots ,0,{\underline {+1,0}},{\underline {-1,0}},0,0\rangle\ + \
|0,\cdots ,0,{\underline {-1,0}},{\underline {-1,0}},0,0\rangle \big ] &
\nonumber\\
&\big [ |0,\cdots ,0,0,0,+1,+1,0,0\rangle\ +\ |0,\cdots
,0,0,0,-1,-1,0,0\rangle\
 +&
\nonumber\\
& |0,\cdots ,0,0,0,+1,-1,0,0\rangle\ + \ |0,\cdots ,0,0,0,-1,+1,0,0\rangle\big
]
 &
\nonumber\\
&\big [ |0,\cdots ,0,+1,+1,0,0,0,0\rangle\  +
\ |0,\cdots ,0,-1,-1,0,0,0,0\rangle +&\nonumber\\
& |0,\cdots ,0,+1,-1,0,0,0,0\rangle\ +
\ |0,\cdots ,0,-1,+1,0,0,0,0\rangle\big ]&\nonumber\\
& {\bf U(1)^2}& \nonumber\\
&\big [ |0,\cdots ,0,0,0,+1,+1,0,0\rangle\ +
\ |0,\cdots ,0,0,0,-1,-1,0,0\rangle\ - &\nonumber\\
& |0,\cdots ,0,0,0,+1,-1,0,0\rangle\ -
\ |0,\cdots ,0,0,0,-1,+1,0,0\rangle\big ] &\nonumber\\
&\big [ |0,\cdots ,0,+1,+1,0,0,0,0\rangle\ +
\ |0,\cdots ,0,-1,-1,0,0,0,0\rangle\ - &\nonumber\\
& |0,\cdots ,0,+1,-1,0,0,0,0\rangle\  -
\ |0,\cdots ,0,-1,+1,0,0,0,0\rangle \big ]&
\end{eqnarray}
where underlining means that all possible permutations must be
properly considered. These are the only states
simultaneously invariant under $\Theta $ and $\Omega $
involving only momenta satisfying $P \cdot L = \ int$.
The above states can be organized into  currents
by checking the operator product expansions (OPEs) that reflect
their corresponding algebras \cite{km}.
Notice that the $SO(10)$
states are orthogonal to those of $SO(8)$ and $U(1)^2$, i.e.
their mixed OPEs are regular. Notice also that the $U(1)$
combinations are chosen so that they are orthogonal to $SO(8)$.
The $SO(8)$ group is realized at level one, since it contains the
level one subgroup $SO(4)$ untouched from the beginning.
On the other hand the $SO(10)$ algebra is realized at level two
as can be verified directly from the OPEs and indirectly in
other ways explained below.

In the untwisted sectors $U_1,U_2$ and $U_3$, the corresponding
left-moving vertices transform under $(\Theta , \Omega )$ with
eigenvalues $(-1,1), (1,-1)$ and $(-1,-1)$ respectively. The momenta
involved must also satisfy $P\cdot L = \ int$.
In sectors $U_1$ and $U_2$ there are matter fields transforming as
$(1,8)$ and with different $U(1)$ charges.
In the $U_3$ sector we find the states
\begin{eqnarray}
& \partial F_I\ , \ I=1,\cdots,10  & \nonumber\\
& |\underline{+1,-1,0,\cdots,0},0,0,0,0,0,0\rangle\
-\ |\underline{-1,+1,0,\cdots,0},0,0,0,0,0,0\rangle &
\end{eqnarray}
These states have no $U(1)^2$ charges and belong to a
$(54,1)+(1,1)$ representation of $SO(10)\times SO(8)$.
Checking the structure of the 54 of $SO(10)$ from OPEs is
cumbersome. Fortunately, there is a simpler argument to support
this fact. Since the orbit states must have $h_{KM}=1$ and they
are neutral under $U(1)^2$ and $SO(8)$, they can only belong to
a 54 that precisely has $h=1$ at $k=2$ as shown in Table \ref{tuno}.
In $U_3$ we also find
\begin{eqnarray}
& \big [ |0,\cdots ,0,{\underline {+1,0}},{\underline {+1,0}},0,0\rangle\  - \
(0,\cdots ,0,{\underline {-1,0}},{\underline {+1,0}},0,0\rangle\ - &\nonumber\\
& \ |0,\cdots ,0,{\underline {+1,0}},{\underline {-1,0}},0,0\rangle\ + \
|0,\cdots ,0,{\underline {-1,0}},{\underline {-1,0}},0,0\rangle \big ]&
\nonumber\\
\end{eqnarray}
These are four singlets, charged under the $U(1)$s
only.

Let us now examine the twisted sectors. Consider first
the sector twisted by $\theta$ and the automorphism $\Theta $.
The left-handed mass formula is given by eq. (\ref{ttwi}). Since,
in this case $E_0+E_{\Theta }=1$ we must have $N_L=0$ and also
$P_T=0$ because $L_T=0$. The quantum numbers of the massless states
are then essentially given by the degeneracy of the vacuum.
The invariant lattice $I$ is the root lattice of $SO(8)$. Its dual
is the weight lattice that has four conjugacy classes.
Hence, $|I^*/I|=4$.
Substituting these value together with $det'(1-\Theta)=2^{12}$
in  eq. (\ref{deg}) we find $D_{\Theta }=32$. The
Wilson line $L$ merely shifts the position of the fixed points
but does not affect the counting.

The value of $D_{\Theta }$ suggests
that this $\theta$ sector contains a $(16,1)+({\overline {16}},1)$.
This guess is confirmed by analyzing the
equivalent model in terms of shifts instead
of automorphisms, {\it before} adding the Wilson line.
In the shift formulation we easily verify that the $\theta$ sector
contains those multiplets. The quantum numbers must be
the same for the equivalent model obtained through automorphisms.
Moreover, they must be the same in the $k=2$ model that is
continuously connected by varying the Wilson line.
The number of the $(16,1)$ and $(\overline{16},1)$ multiplets
depends on the specific form of the $Z_2\times Z_2$ rotations
$\theta$ and $\omega$ as explained in the Appendix. With the choice
leading to the multiplicity factor in eq. (\ref {puno}) we obtain three
$SO(10)$ generations plus one antigeneration.
The $\omega$ sector also gives three $(16,1)$ and one
$(\overline{16},1)$ with different $U(1)$ charges. In the
$\theta\omega$ sector we obtain states transforming as $(10,1)$,
$(1,8)$ and singlets.

Altogether the spectrum of this GUT model is given in
Table \ref{ttres}. The charge $Q$
is non-anomalous whereas $Q_A$ is anomalous.
The gravitational, cubic and mixed gauge anomalies of $Q_A$ must
be in the correct ratios in order to be cancelled
by the 4-D version of the Green-Schwarz mechanism
\cite{gs} .
In particular, the mixed anomalies of $Q_A$ with $SO(10)$
and $SO(8)$ should be in the same ratio as the levels $k_{10}/k_8=2$.
We find $Tr  Q_A / Tr  Q_A^3 = 24/3$; ${\cal B}_8 / Tr  Q_A^3 = 1/3$
and ${\cal B}_{10} / Tr Q_A^3 = 2/3$, where ${\cal B}$ is the mixed
anomaly coefficient. These expected results furnish a consistency
check of our construction.

\begin{table}
\begin{center}
\begin{tabular}{|c|c|c|c|}
\hline
$Sector $
& $SO(10)\times SO(8)$ & $Q$ &  $Q_A$   \\
\hline
$  U_1  $  &     (1,8) &   1/2   &   1/2  \\
\hline
& (1,8)   &    -1/2   &  -1/2  \\
\hline
$  U_2  $ & (1,8)   & -1/2 & 1/2  \\
\hline
& (1,8)    &   1/2  &  -1/2  \\
\hline
$   U_3  $ & (54,1)   & 0   &  0  \\
\hline
& (1,1)    &    0  & 0   \\
\hline
& (1,1)    &  0  & 1    \\
\hline
& (1,1)    &   1  & 0   \\
\hline
& (1,1)    &  -1  & 0   \\
\hline
& (1,1)    &  0  & -1   \\
\hline
$\theta$   & $3(16,1)$   &  1/4  &  1/4  \\
\hline
& $(\overline{16},1)$   &  -1/4  &  -1/4  \\
\hline
$\omega$   & $3(16,1)$   &  -1/4  &  1/4  \\
\hline
& $(\overline{16},1)$   &  1/4  &  -1/4  \\
\hline
$\theta\omega$ &    $ 4(10,1)$   & 0   & 1/2  \\
\hline
& $4(10,1)$   & 0   & -1/2  \\
\hline
& $3(1,8)$   & 0   & 1/2  \\
\hline
& $(1,8)$   & 0   & -1/2  \\
\hline
& $8(1,1)$   & 1/2   & 0  \\
\hline
& $8(1,1)$   & -1/2   & 0  \\
\hline
\end{tabular}
\end{center}
\caption{Particle content and charges of Example 1.}
\label{ttres}
\end{table}

We now wish to discuss an important feature of the GUT Higgs and
its singlet partner appearing in the $U_3$ sector. In the 0-picture
the full emission vertex operator for the singlet has the form
\begin{equation}
\partial X_3\ \otimes \ \sum _{I=1}^{10}\partial F_I
\label{svec}
\end{equation}
A Vev for this field precisely corresponds to the Wilson line
background $L$ in eq. (\ref{wil}).
The fact that this background may be varied continuously
means that this singlet is a
{\it string modulus}, a chiral field whose scalar potential is
flat to all orders. Indeed, using the
discrete $Z_2$ R-symmetries of the right-handed sector, it can
be proven that its self-interactions vanish identically.

The GUT Higgs contains the other 9 linear combinations of
$\partial F_I$. These give the diagonal elements
of the symmetric traceless matrix chosen to represent the 54-plet.
the associated vertex operator is
\begin{equation}
\partial X_3\ \otimes \ \sum _{I=1}^{10} c_I\partial F_I
\ \ ;\ \ c_I\in {\bf R}, \  \ \sum_Ic_I=0 \ .
\label{hvec}
\end{equation}
Vevs for these nine components of the 54 would correspond to the
presence of more general Wilson backgrounds of the form
$L=(\lambda_1,\lambda_2,\cdots ,\lambda_{10},0,0,0,0,0,0)$ with
$\sum _{I=1}^{10} \lambda_I=0$.
These more general backgounds break the symmetry further to some
$SO(10)$ subgroup like $SU(4)\times SU(2)_L\times SU(2)_R$.
The fact that these other nine modes may
be continuously varied means that they are also string
moduli or, more generally, that the 54-plet of $SO(10)$ in this
model is itself a string modulus! We find that this property of the
GUT-Higgs behaving as a string modulus, on equal footing
with the compactifying moduli $T_i$, is very remarkable.

We have constructed with simple methods a level $k=2$
$SO(10)$ GUT model with a single GUT-Higgs transforming as a 54.
This model has other interesting properties, particularly in the couplings
of the Higgs sector as well as in the one-loop Fayet-Illiopoulos
(see section 7).

The example summarized in Table \ref{ttres} belongs to a whole
class of models obtained through continuous Wilson lines.
A general characteristic is that they are $SO(10)$ models
in which the GUT Higgs is a 54 multiplet.
Moreover, there is only one such GUT Higgs coming from
the untwisted sector and behaving like a string modulus.
On the other hand, the rest of the particle content is model
dependent. This includes the number of generations,
existence of Higgses 10s, $(16+{\overline {16}})$s,
hidden gauge group, etc.. For instance, the number of generations
can be changed by adding discrete Wilson lines to the original \
orbifold. There are no $SO(10)$ models in this class
with 45s of Higgses instead of 54s.
Although our search has been far from complete, we have not
found $SU(5)$ models in this class.
For reference we will now give two more examples of this class
skipping the details. They could guide the
reader in looking for different models.

Our second example
is also based on the $Z_2\times Z_2$ orbifold but this time
one of the $Z_2$s is realized by a reflection and the other by a
shift. The actual embedding is
\begin{eqnarray}
\theta\ :& \ \Theta (F_1,F_2,\cdots ,F_{16})\ =\
(-F_1,-F_2,\cdots ,-F_{11},-F_{12},F_{13},F_{14},F_{15},F_{16})&
\nonumber  \\
\omega\ :  &  B=(0,0,0,0,0,0,0,0,0,0,
{1\over 2},{1\over 2},{1\over 2},{1\over 2},{1\over 2},{1\over 2}) &
\label{seg}
\end{eqnarray}
At this stage the gauge symmetry is $SO(10)^3\times U(1)_A$.

After turning on a Wilson line of the form in eq. (\ref{wil}),
the gauge group breaks to $SO(10)\times SO(10)\times U(1)_A$ with
the first $SO(10)$ realized at level $k=2$.
There are no matter fields in the $U_2$ and $U_3$ sectors whereas
$U_1$ contains $(54+1,1)$ + $2(1,10)$.

In both $\theta$ and $\omega$ sectors there are
3 copies of $(16,1)$ and one $(\overline{16},1)$. In the $\theta\omega$ sector
there are instead
3$(1,{\overline {16}})$ and one $(16,1)$.
The initial level one orbifold in this example may equivalently be
constructed  through the five-fold embedding discussed
in section 3. This model is further discussed in section 6 where it is
constructed using a different method.

Our third example is based in the symmetric $Z_4$ orbifold.
The single generator $\theta$ with shift
$v=({1\over 4},{1\over 4}-{1\over 2})$ is
realized through the automorphism given by
\begin{equation}
v\ : \ \Theta (F_1,F_2,\cdots ,F_{16})\ =\
(F_2,-F_1,F_4,-F_3,-F_5,-F_6,\cdots ,-F_{15},-F_{16})
\label{taut}
\end{equation}
After adding a Wilson-line of the form
$L=(0,\cdots, 0, \lambda , \lambda  ,\lambda ,\lambda , \lambda ,
\lambda  )$,
we are left with gauge group $SO(10)\times SU(2)^5\times U(1)$
with the $SO(10)$ realized at $k=2$. This model
again has a 54 in the untwisted sector, no massless states in the
$\theta$-twisted sector and $10$s plus hidden matter
in the $\theta^2$ sector.

\section{Constructing level two string models through
permutation modding}

In this second method
the basic observation is that, when two identical gauge factors
$G\times G$ belonging to a starting level $k=1$ model are exchanged,
the diagonal group $G_D$ at level $k=2$ emerges as the
survivor of the projection over permutation invariant states
\cite{fiq} .
This general idea may be implemented in orbifold constructions,
in essentially two different ways :

{\bf i.} The order-two permutation is associated to  one of the
twists defining the orbifold. This means that
in a $Z_2\times Z_N$ orbifold, $Z_2$ is embedded through the
a permutation $\Pi $ of two identical gauge factors.
The $Z_N$ action is realized in the usual way, through
a shift $V$ in the lattice $\Lambda _{16}$.
Since we are dealing with Abelian orbifolds, the two
operations must commute. This implies the constraint
\begin{equation}
V\ =\ \Pi V\ \  mod  \ \Lambda _{16}
\label{p1}
\end{equation}
where $\Pi V$ is the shift obtained upon permutation.
This turns out to be a very strong requirement. A more general
situation in $Z_M \times Z_N$, with $M$ even, may be imagined
by embedding the $Z_M$ twist as an order two permutation
plus a shift in the $\Lambda _{16}$ lattice. Nevertheless,
eq. (\ref {p1}) must still be satisfied.

Additional requirements come from modular invariance (level-matching)
which essentially limit the number of pairs of coordinates which may be
permuted. This is discussed below.

{\bf ii.}
The order two permutation $\Pi $ mimics the effect of a
{\it quantized} Wilson line in the orbifold. Since
this Wilson line has order two,  the
original $Z_N$ or $Z_N\times Z_M$ orbifold must be of even order.
 Thus $Z_3$, $Z_3\times Z_3$ and $Z_7$ cannot be used for this purpose.

Consistently embedding the space group into the gauge degrees of
freedom imposes again severe constraints. Interestingly enough,
these constraints depend on the way in which the $Z_N$ action
is realized on the six dimensional lattice.
In order to exemplify this point,
let us consider a $Z_N$ orbifold, defined
through a given twist $\theta $ with associated gauge lattice shift
$V$.  The corresponding space-gauge group twisting element is denoted
$(\theta , 0| 1, V)$.
We also add a discrete Wilson line along, say, the
compactifying lattice vector $e_1$. The associated group element
$(1,e_1|\Pi , W)$  implements  a shift $e_1$ in the compactifying
lattice and simultaneously acts as permutation $\Pi $ plus
a shift $W$ in the gauge lattice.

The product element $(1,e_1|\Pi , W)(\theta , 0|1,V)=
(\theta , e_1|\Pi , \Pi V+W)$,
must belong to the space-gauge twisting group. By applying this
element $N =2j$ times we get
\begin{eqnarray}
(\theta , e_1 |\Pi, \Pi V+W)^N   =  \nonumber  \\
(\theta ^N , e_1+ \theta e_1+...+\theta ^{N-1}e_1   |  \Pi ^N,
\Pi V+W+\Pi (\Pi V +W)+...& +\Pi ^{N-1}(\Pi V+W))  =  \nonumber \\
 (1,0|1, j[\Pi V+W]+j[V + \Pi W]) & & \nonumber
\label{p2}
\end{eqnarray}
 For the embedding to be a consistent homomorphism of the
space group into the gauge degrees of freedom, the above element
should be trivial. Therefore,
\begin{equation}
j[\Pi (V+W) +(V+W)]  \in \ \Lambda _{16}
\label{p3}
\end{equation}

This is a necessary constraint, but depending on the compactifying lattice
there could even be additional ones.
Let us consider, as an example, the  $Z_4$ orbifold defined by the eigenvalues
$1/4(1,1,-2)$. On a $SU(4) \times SU(4)$ lattice, above equation
with $j=2$ is also sufficient.
In particular, this means that order four Wilson lines are admitted.
For the cubic $SO(4)^3$ lattice there is another constraint
due to the relation
\begin{equation}
\label{p4}
e_1 + \theta ^2e_1=0
\end{equation}
that implies
\begin{equation}
\left.
\begin{array}{c}
\Pi W+ W  \\
2[\Pi V +V]
\end{array}  \right  \}
\in \ \Lambda _{16}
\end{equation}
    Only order two Wilson lines are allowed in this case. Moreover, if we
chose to associate the permutation Wilson line to the third $SO(4)$ lattice,
the even more severe constraint  equation (\ref {p1}) is found.

{}From this example we learn that the constraints,
and therefore the model building possibilities,
coming from the embedding of
the space group into the gauge degrees of freedom, crucially depend on the
compactifying lattice chosen.
This is due to the existence of relations among twist
and lattice vectors. A similar situation arises when, in a
given lattice, there exist different inequivalent twists
that can realize the orbifold action as happens in
$Z_2 \times Z_2$ examples in Appendix.

The construction  of type {\bf i.} and {\bf ii.}  models
follows the usual rules of orbifold model building.
Some distinguishing features appear in those twisted sectors in
which the twist in the compactified dimensions is accompanied by
a permutation in the gauge degrees of freedom.
We will now discuss this kind of sectors and provide some examples
to illustrate the whole procedure.

Consider an initial $k=1$ model including a group $G \times G$,
with  $G$ of rank $R$ ($2R\leq 16$).
Denote the Cartan generators of the first (second) $G$ factor
by $\partial X$($\partial Y$) and the remaining (up to 16) by
$\partial Z$. In the twisted sectors, taking into account the
permutation modding, we have the following
boundary conditions for the three types of gauge coordinates
\begin{eqnarray}
 X(u+\pi )\ & =\ Y(u)\ + \pi P_1\ +\ \pi V_1   & \nonumber\\
 Y(u+\pi )\ & =\ X(u)\ + \pi P_2\ +\ \pi V_2   & \nonumber\\
 Z(u+\pi )\ & =\ Z(u)\ + \pi P_3\ +\ \pi V_3   &
\label{p5}
\end{eqnarray}
where $u=\sigma -\tau $ is the left-handed world-sheet variable,
 $P_i$ are components of vectors $P\in \Lambda _{16}$ and
$V_i$ are components of a shift which might be
present in the specific twisted sector considered.
We can write mode expansions for the coordinates $X$ and $Y$
corresponding to the two gauge factors,
\begin{eqnarray}
  X(u)\ &=\  X_0\ +\ M_1u\ +\ {i\over 2}\sum{{x_r}\over r}e^{-2iru} &
\nonumber\\
  Y(u)\ &=\  Y_0\ +\ M_2u\ +\ {i\over 2}\sum{{y_r}\over r}e^{-2iru} &
\label{p6}
\end{eqnarray}
where $M_1$ and $M_2$ are the quantized momenta.
When boundary conditions (\ref{p5}) are imposed it follows that
\begin{eqnarray}
 M_1\  =\ M_2\ =\ M\ =\ {{(P+V)}\over 2}& & \\
 x_r\ =\ e^{2i\pi r}y_r \ \ ;\  \ y_r\ = \ e^{2i\pi r}x_r  & &
\label{p7}
\end{eqnarray}
The second conditions in this equation indicate  that $r=m$ or
$r=m+1/2$. Therefore, gauge oscillator numbers are either
integer or semi-integer.
Permutation modding contributes
to the vacuum energy by increasing it by $R/16$, where $R$ is the
number of permuted pairs of coordinates.
 Altogether, we conclude that the left-handed mass formula in a
permuted twisted sector is given by
\begin{equation}
{1\over 8}m_L^2\ =\ N_L\ +\ {{(P_{\pi } \ +V_{\pi })^2}\over 4}\ +\
{{(P_3+V_3)^2}\over 2}\ +\ E_0\
+\ {R\over {16}}\ -1
\label{p8}
\end{equation}
 where $P_{\pi }+V_{\pi }=P_1+P_2+V_1+V_2$ = $2M$. Here $N_L$ stands
for the oscillator numbers  from both the
compactifying twist and the permutation modding and $E_0$ is the
vacuum energy from the twist, as given in Table \ref{tuno}.
This is  what we essentially need to compute the
massless states in these permuted sectors.

As mentioned above, for the permutation method to really give rise
to a level 2 model,
the permutation modding must be performed among gauge coordinates
$X$ and $Y$ which {\it do not} belong to the same gauge group.
Instead they must correspond to gauge factors $G_1$ and $G_2$
well differentiated.

Since we want to obtain GUT gauge groups such as
$SU(5)$, $SO(10)$ and even $E_6$, a natural possibility is to
embed each of the two identical gauge  groups into a different
$E_8$ factor of the $E_8\times E_8$ heterotic string and then do
a permutation modding of the coordinates of both $E_8$s. Let us
assume that the permutation is associated to a twist, as considered
in {\bf i.}. Therefore eq. (\ref {p1}) must be verified, leading to
$E_8\leftrightarrow E_8'$
symmetric shifts of the form
\begin{equation}
V\ =\ {1\over N}(d_1,d_2,...,d_8)\otimes {1\over N}(d_1,d_2,...,d_8)
\label{p9}
\end{equation}
associated to the 6-dimensional twist $v={1\over N}(a,b,c)$.
It is easy to prove, however, that there are no such
symmetric shifts which are
modular invariant for any {\it even order} $Z_N$ orbifold.
Indeed, the modular invariance condition eq. (\ref{Vcond}) implies
\begin{equation}
2(d_1^2+...+d_8^2)\ -\ (a^2+b^2+c^2)\ =\ 0\ \ mod\ 2N
\label{p10}
\end{equation}
but the first term is necessarily $0 \ mod \ 4$, whereas for all
Abelian orbifolds, $(a^2+b^2+c^2)$  is always $2 \ mod \ 4$.
Since $N$ is even by hypothesis, we conclude
thet (\ref{p10}) cannot be fulfilled.

The second type of construction  leads to similar conclusions if
eq. (\ref {p2}) must be satisfied.
Thus we see that the permutation
modding mechanism in $E_8\times E_8$ through symmetric orbifolds
cannot possibly work.

Alternatively, we can try to start with models constructed from
$Spin(32)/Z_2$ such as the
$SU(5)^3$ and $SO(10)^3$ models obtained in section 3 by using the
five-fold embedding.
Again, the requirements on the gauge shifts
and level matching conditions are usually very restrictive.
For example, it can be proved with complete generality that,
in the $Z_2\times Z_2$ orbifold, conditions of type
(\ref {p3}) ($j=1$ in this case)
and level-matching can only be consistent when
either 4 or 8 pairs of coordinates are permuted.
This is valid even if the permutation is accompanied
by an arbitrary gauge shift. In order to obtain
an $SU(5)$ or an $SO(10)$ GUT, the modding of 5
pairs of coordinates is necessary. Hence, this realization is ruled out.
Moreover, this proof may be extended to all those even orbifolds where
the $Z_2$ sector of the orbifold feels the Wilson line permutation.

A way out to this limitation may be found in some cases.
For example, this restriction may be avoided in a $Z_4$ orbifold realized,
either by
using Coxeter rotations on the $SU(4)^2$ root lattice
or by assigning the permutation
line to a third lattice direction in $SO(4)^3$.
In both cases the $\theta ^2$ sector
does not split in the presence of an order two Wilson line (see Appendix ).
However, the second situation is excluded in practice, due to
restriction (\ref {p1}).

Let us discuss a particular example.
We start with the $Z_4$ orbifold in the $SU(4)^2$ lattice and
embedding
\begin{equation}
V  =  ({1\over 2},{1\over 2},{1\over 2},{1\over 2},
{1\over 2},0,0,0,0,0,0,{1\over 4},{1\over 4},0,0,0)
\end{equation}
In order to lower the number of generations we also turn on
Wilson lines $L_1=L_2=L_3=L$ with the specific order four $L$
\begin{equation}
L  =  (0,0,0,0,0,0,0,0,0,0,0,{1\over 4},-{1\over {4}},{1\over 4},
{1\over 4},{1\over 2})
\end{equation}
The emerging gauge group is
$SO(10)\times SO(12) \times SU(2) \times U(1)^4$. The massless
spectrum is found using the projectors described in the Appendix.
The results are shown in Table \ref{tp1}.

\begin{table}
\begin{center}
\begin{tabular}{|c|c|c|c|c|c|}
\hline
$Sector$
   & $SO(10)\times SO(12 )\times SU(2)$ & $Q_1$ & $Q_2$ & $Q_3$ & $Q_4$ \\
\hline
$  U_1, U_2 $  & 2 (1,1,2)          &  1     &   0   &  -1   &    0  \\
\hline
	   &    2 (1,1,2)           &  0     &   1   &   1   &    0  \\
\hline
$  U_3$    &      (1,1,1)           &  1     &   1   &   0   &    0  \\
\hline
	   &      (1,1,1)           & -1     &  - 1   &   0   &    0  \\
\hline
	   &      (10,12,1)         &  0     &   0   &  0   &    0  \\
\hline
$ \theta   $    &    4 ($\bar {16}$,1,1)     & 1/4     &  1/4 &   0   &    0
\\
\hline
$ (\theta ^2, 2V) $    &    2 (1,12,1)       & 1/2     &  1/2 &   0   &    0
\\
\hline
	   &    2 (10,1,1)       &-1/2     & -1/2 &   0   &   0  \\
\hline
	   &    2 (1,1,1)        & 1/2      & 1/2   &  0   & $ \pm 1$  \\
\hline
	   &     (1,1,2)         & 1/2     &  1/2 & $ \pm 1$     &    0  \\
\hline
	   &     (1,1,2)         & -1/2       & -1/2 &  $ \pm 1$    &   0  \\
\hline
$Osc.$     &     (1,1,1)         & -1/2      & 1/2  &   0     &   0  \\
\hline
	   &    (1,1,1)          & 1/2       & -1/2  &   0     &   0  \\
\hline
$ (\theta ^2, 2V+2L)  $  &     (1,12,1)         & 0     &  0 &  $ \pm 1$    &
 0  \\
\hline
	   &     (10,1,1)          & 0     & 0  &   $ \pm 1$  &   0  \\
\hline
	   &    2 (1,1,1)          & 0     &  $ \pm 1$ &   -1   &    0  \\
\hline
	   &     2 (1,1,1)         & $ \pm 1$    & 0   &   1   &    0  \\
\hline
	   &     (1,1,1)           & 0     &  0 &  $ \pm 1$   &  $ \pm 1$  \\
\hline
$Osc.$     &    2(1,1,2)           & 0       & 0 &   0     &   0  \\
\hline
\end{tabular}
\end{center}
\caption{Particle content and charges before modding by
 permutations.}
\label{tp1}
\end{table}

The sector $\theta$ is split into four sub-sectors but
we do not find any massless states in those with shift
$(V+nL)$, $n=1,2,3$. Massless generations are only found
in the $(\theta,0)$ sub-sector. In this way the Wilson
line effectively reduces the number of generations.
State multiplicities are determined using the projectors
discussed in the Appendix.

Permutation modding of the first $SO(10)$ factor with the $SO(10)$ subgroup
of $SO(12)$
\begin{eqnarray}
 \Pi (F_1,F_2,F_3,F_4,F_5,F_6,F_7,F_8,F_9,F_{10},
 F_{11}, \cdots ,F_{16}) & = &  \nonumber \\
 (F_6,F_7,F_8,F_9,F_{10},F_1,F_2,F_3,F_4,F_5,F_{11}, \cdots ,F_{16}) & &
\label{eqn:16}
\end{eqnarray}
may now be included as a Wilson line in the second
$SU(4)$ lattice.
The surviving gauge group is therefore $SO(10)_D \times SU(2) \times U(1)^5$
and matter states will organize into its corresponding representations.
For example in the untwisted sector we obtain,
$$2(1,2)+2(1,2)+(1,1)+(1,1)+(1,1)+(54,1)$$
 where $SO(10)$ singlets are split
according to the different $U(1)$ charges.
The $54$ of $SO(10)$
is found (see discussion in section 4)
when non invariant states are projected out  from the starting
$(10,12,1)$ representation.

The twisted sectors in the initial model
in Table \ref{tp1} will split into sub-sectors that may or may not
detect this second order permutation Wilson line. For example,
the $\theta $ sector includes a $V$ sub-sector,
corresponding to  fixed points $(0,0)$ and $(0,w_2)$, not
feeling the permutation line. There is also a $V_{\Pi}$ sub-sector,
corresponding to  fixed points $(0,w_1)$ and $(0,w_3)$, now feeling
$\Pi$. Both sub-sectors contribute with a $2(\overline{16}, 1)$
representation and therefore we end up with a four generation $SO(10)$
model at level $k=2$. Sixteen generations are found if the
Wilson line $L$ is not present.

As another example, let us mention that the third model of section 4,
built up through the automorphism (\ref {taut}) plus the addition of a
continuous Wilson line, may be reobtained by permutation modding.
In fact, this is achieved by considering the  shift
$V=\frac{1}4 (1,1,2,2,2,2,2,2,2,0,0,0,0,0,0,0)$ which leads  to an
$SO(14)\times SO(14)\times U(1)$ gauge group at level 1, and then adding
a permutation Wilson line which exchanges the $SO(10)$'s contained in each
$SO(14)$.

As the above examples show, the explicit models found are quite similar to
the ones obtained through continuous Wilson lines. However not all the
models obtained through permutation modding should be obtainable through
the first method because the permutation modding involves a discrete
projection on permutation-invariant states.

\section{Constructing level two models from Higgs mechanism through flat
directions}

In Refs. \cite{imnq,finq} it was remarked that the mechanism
of gauge symmetry breaking through continuous Wilson lines may be
understood perturbatively in terms of flat directions in the
scalar potential of massless charged {\it untwisted } fields.
In \cite{fiq} it was shown that in fact there are flat directions
which continuously connect level one to higher level string theories
and an explicit $k=3$ example was constructed. More generally,
higher level theories may be obtained by giving vevs along flat
directions to {\it both twisted and untwisted} massless scalars
\cite{finq} .
The general procedure uses 4-D supersymmetry
in the effective field theory to impose the
flatness conditions
\begin{eqnarray}
& \langle W\rangle=0  \ \ \ ;  \ \ \
\langle F_i\rangle=\langle \frac{\partial W}{\partial \phi _i}\rangle=0 &
\nonumber \\
& \langle D_{\alpha }\rangle=
\langle g_{\alpha }\phi _i^*(T_{\alpha })^i_j\phi ^j\rangle=0 &
\label{eqn:36}
\end{eqnarray}
where $W$ is the superpotential, $\phi _i$ the scalar fields
and $g_{\alpha }$ and $T_{\alpha }$  the couplings and
 generators of the gauge group.

In the presence of an
anomalous $U(1)_A$ whose anomaly is canceled through a
Green-Schwarz mechanism
\cite{gs} , there is a one-loop modification
to the $D$-term. This is a dilaton-dependent Fayet-Illiopoulos
term which has to be
added \cite{dterm}
\begin{equation}
D_A\ =\ \sum _iq_A^i|\phi _i|^2\ +\
{g\over { 192 \pi^2 \sqrt{k_A}} } \, TrQ_A
\label{eqn:37}
\end{equation}
where $TrQ_A$ is the trace of the anomalous $U(1)_A$ over the
complete massless spectrum, $g$ is the gauge coupling constant
and $k_A$ is the normalization (``level") of the $U(1)_A$.
In the following discussion we obviously assume
that the value of $g$, determined by the dilaton vev, has been
fixed by some non-perturbative dynamics that will not be discussed here.

Notice that for the usual classical vacuum
$\langle \phi _i \rangle=0$ this
extra term would induce supersymmetry breaking because
$\langle D_A \rangle\not= 0$. However,
what normally happens is that some of the $\phi _i$s
are forced to have a vev and cancel the one-loop piece. For this
to happen it is crucial that there  exist fields $\phi _i$ in
the massless spectrum with charge $q_i$ of sign {\it opposite}
to that of $TrQ_A$. Although there is no general principle
that guarantees the existence of such fields,  the fact is that
up to now a 4-D string in which this is not the case has not been found.
Thus, in the presence of an anomalous $U(1)_A$, classical string vacua
are generically unstable but there is typically a nearby minimum
which constitutes a one-loop stable vacuum. As we shall show,
it turns out that the Fayet-Illiopoulos term often plays an important
role in the construction of our class of GUT models.

For our particular interest of building GUTs, we start with
level one models with gauge group and massless chiral fields
of the type
\begin{eqnarray}
& SU(5)\times SU(5)\times G& \ ;\ \ (5,{\bar 5})\ , ({\bar 5},5)
\nonumber \\
& SO(10)\times SO(10)\times G' &\ ; \ \ (10, 10)\
\label{eqn:38}
\end{eqnarray}
 Giving appropriate vevs with vanishing D-terms
to the chiral fields, the duplicated groups
are spontaneously broken to the diagonal subgroups $SU(5)_D$,
$SO(10)_D$  which are realized at level two. Of course, it must also
be checked
that the F-terms also vanish, which is sometimes non-trivial.
The fact that the level is increased to two is explainable since one
knows that, when a group $G^M$ is broken to $G_{diag}$, the
coupling constant must be rescaled as $g\rightarrow g/\sqrt{M}$.
In the string context this means that the original level
is rescaled as $k\rightarrow kM$
\cite{fiq} .

Starting with duplicated
groups is not the only possibility, one can also start with a level
one model with group e.g. $SU(5)\times G$ where $G\supseteq SU(5)$ and
similarly for $SO(10)$. The first model discussed in
section 4 may be understood as an example of this type since it
starts with gauge group $SO(10)\times SO(18)\times U(1)^2$,
before adding the continuous Wilson line.
By giving appropriate vevs
to a field $(10,18)$ present in the untwisted sector, the final theory
is the level $k=2$ $SO(10)$ model displayed in Table \ref{ttres} .
To simplify the
discussion though we will focus on models with repeated
gauge group factors.

We already mentioned a class of orbifold models that naturally leads to
replication of gauge groups $SU(5)$ and $SO(10)$, namely
the models obtained through the five-fold embedding we discussed
in section 3. Let us then examine some of these.

{\it Flat direction model I (FD-I)}

Consider the simplest five-fold embedding example based on
the $Z_2\times Z_2$ orbifold. It may be equivalently defined through
the embedding
 \begin{eqnarray}
A & = & ({1\over 2},{1\over 2},{1\over 2},{1\over 2},{1\over 2},
{1\over 2},0,0,0,0,0,0,0,0,0,0) \nonumber\\
B & = & ({1\over 2},0,0,0,0,0,{1\over 2},{1\over 2},{1\over 2},
{1\over 2}, {1\over2},0,0,0,0,0)
\label{eqn:39}
\end{eqnarray}
One can easily compute the massless spectrum  with the help of the
generalized GSO projector given in the Appendix. The
multiplicities in the twisted sectors depend on the particular
realization chosen for the $Z_2 \times Z_2$ twists acting on the cubic
lattice. With the choice leading to the projectors eq. (\ref{abcpro}),
we obtain the results displayed in Table \ref{tcinco} under the title
$k=1$.

 \begin{table}
\begin{center}
\begin{tabular}{|c|c|c|c|c|}
\hline
$Sector $
   & $k=1\ :SO(10)\times SO(10)\times SO(10)$ & $Q_A$ &
 $SO(10)_{k=2}\times SO(10)_{k=1} $    &   $Q_A$   \\
\hline
$  U_1  $  &     $(1,10,1)$ &   +1   &      &  \\
\hline
       &
 $(1,10,1)$   &    -1   &     &  \\
\hline
       &
 $(10,1,10)$   &    0   &   &  \\
\hline
$  U_2  $  &     $(10,1,1)$ &   +1   &      & \\
\hline
       &
 $(10,1,1)$   &    -1   &     &  \\
\hline
       &
 $(1,10,10)$   &    0   &     &  \\
\hline
$  U_3  $  &     $(1,1,10)$ &   +1   &   $(1,10)$   &   +1  \\
\hline
       &
 $(1,1,10)$   &    -1   & $(1,10)$    &  -1  \\
\

       &
 $(10,10,1)$   &    0   & $(54,1)+(1,1)$    &  0  \\
\hline
$\theta$    &
$3(16,1,1)$   &  -1/2  &   $3(16,1)$  &  -1/2  \\
\hline
     &
$({\overline {16}},1,1)$   &  +1/2  &   $({\overline {16}},1)$  & +1/2  \\
\hline
   $\omega$  &
$3(1,16,1)$   &
   -1/2    & $3(16,1)$    &   -1/2  \\
\hline
     &
$(1,{\overline {16}},1)$   &  +1/2  &  $({\overline {16}},1)$  & +1/2  \\
\hline
$\theta\omega$ &  $3(1,1,{\overline {16}})$   & -1/2
& $3(1,{\overline {16}})$   &   -1/2  \\
\hline
     &
$(1,16)$   &  +1/2  &   $(1,16)$  &  +1/2  \\
\hline
\end{tabular}
\end{center}
\caption{Particle content and charges of the model FD-I, before
and after taking the flat direction.}
\label{tcinco}
\end{table}

Notice that in the $U_3$ sector there is a $(10,10,1)$ multiplet whose
field we denote by $\phi ^{ij}$. A vev
$\phi ^{ij}=V\delta ^{ij}$ has vanishing D-term
and breaks the symmetry to the diagonal $SO(10)$. It is also easy to
prove that this field direction is F-flat. Indeed, the discrete
right-moving $Z_2$ R-symmetries or ``H-momentum"
selection rules \cite{finq,fiqs},
forbid any self-coupling of untwisted fields. The renormalizable
Yukawas in the $Z_2\times Z_2$ are of the form
$(U_1\cdot U_2 \cdot U_3)$, $(U_1\cdot \omega \cdot \omega)$,
$(U_2\cdot \theta \cdot \theta)$,
$(U_3 \cdot \theta\omega \cdot \theta\omega)$, and
$(\theta \cdot \omega \cdot \theta\omega)$.
The resulting level two
GUT model  particle content is shown in the right part of the
table and in fact corresponds to an alternative construction of
the second model discussed in section 4. The model has an anomalous
$U(1)$ and an associated dilaton-dependent Fayet-Illiopoulos term.
lt can be made one loop stable by giving a vev to the field $(1,10)$ with
charge $q_A=+1$. This breaks the level one $SO(10)$ group but
does not affect the level two $SO(10)$.
Altogether the GUT has four $16$ generations and appropriate
Higgs fields to break $SO(10)$ down to the SM. Although
apparently the possible Higgs 10-plets get mass along
the flat directions, residual light Higgs doublets
may result for particular values of the $(10,10,1)$ and
$(1,1,10)$ (see the discussion about the doublet-triplet
splitting problem in the next section).

{\it Flat direction model II (FD-II)}

A $SU(5)$ model may be obtained from the previous one
by adding a discrete Wilson line $L$. Specifically, we add
$L_1=L_2=L$ where
\begin{equation}
L\ =\ {1\over
4}(3,1,1,1,1,1,1,1,1,1,1,1,1,1,1,3)
\label{wil5}
\end{equation}
The group is broken to $SU(5)^3\times U(1)^3\times
U(1)_A$ and the spectrum is modified substantially. As explained
in the Appendix, the Wilson line is only detected in the
$\theta$-twisted sector.
The complete chiral spectrum of the $k=1$ model is shown in
Table \ref{tseis}.

\begin{table}
\begin{center}
\begin{tabular}{|c|c|c|c|c|c|}
\hline
$Sector $
   & $SU(5)\times SU(5)\times SU(5)$ & $Q_A$   &  $Q_1$
&   $Q_2 $    &   $Q_3$   \\
\hline
$  U_1  $  &   $(1,5,1)$ &   1   &    0  &   1  & 0  \\
\hline
      &   $(1,{\overline 5},1)$   &    -1   &  0   &  -1  & 0  \\
\hline
      &   $(5,1, {\overline 5})$ & 0   &    1  &   0  & -1 \\
\hline
    &  $({\overline 5},1, 5)$   &  0   &  -1 &  0  & 1  \\
\hline
$  U_2  $  &   $(5,1,1)$ &   1   &    1  &   0  & 0  \\
\hline
    &   $({\overline 5},1,1)$   & -1   &  -1   &  0  & 0  \\
\hline
    &   $(1,5, {\overline 5})$ &   0   &    0  &   1  & -1  \\
\hline
    &   $(1,{\overline 5}, 5)$   & 0   &  0   &  -1  & 1  \\
\hline
$  U_3  $  &   $(1,1,5)$ &   1   &    0  &   0  & 1  \\
\hline
    &  $(1,1,{\overline 5})$   &   -1   &  0   &  0  &  -1  \\
\hline
      &   $(5, {\overline 5},1)$ &   0   &    1  &   -1  & 0  \\
\hline
    &  $({\overline 5}, 5,1)$   &    0   &  -1   &  1  &  0  \\
\hline
$(\theta,0)$    &
$2({\overline 5},1,1)$   &  -1/2  & 3/2    &  0  &0  \\
\hline
    &   $2(10,1,1)$  &  -1/2  & -1/2 & 0  & 0  \\
\hline
    &   $2(1,1,1)$  &  -1/2  & -5/2  &  0  & 0  \\
\hline
$\omega$    &
$3(1,{\overline 5},1)$   &  -1/2  &  0   & 3/2   &  0  \\
\hline
    &   $3(1,10,1)$  &  -1/2  &  0 & -1/2   &  0  \\
\hline
    &   $3(1,1,1)$  &  -1/2  &  0  & -5/2     &  0  \\
\hline
    &   $(1, 5,1)$   &  1/2  &  0   & -3/2  &   0  \\
\hline
    &   $(1,{\overline {10}},1)$  &  1/2   &  0   &  1/2 & 0  \\
\hline
    &   $(1,1,1)$  &  1/2  &  0  & 5/2     &  0  \\
\hline
$(\theta,L)$    &
$(1,1,{\overline 5})$   &  1/4    &  -5/4   & 5/4    &  1/4  \\
\hline
    &   $(1,5,1)$  &  -1/4  & 5/4   & -1/4  & -5/4  \\
\hline
    &   $({\overline 5},1,1)$  &  -1/4  & 1/4  &  -5/4  & -5/4  \\
\hline
&   $(1,1,1)$  &  -3/4  & -5/4   &  5/4    & 5/4  \\
\hline
    & $(1,1, 5)$   &  -1/4    &  5/4   & -5/4    &  -1/4  \\
\hline
    &   $(1,{\overline 5},1)$  &  1/4  & -5/4   & 1/4  & 5/4  \\
\hline
    &   $( 5,1,1)$  &  1/4  & -1/4  & 5/4   & 5/4  \\
\hline
&   $(1,1,1)$  &  3/4  & 5/4   &  -5/4    &  -5/4  \\
\hline
$\theta\omega$ &
$(1,1,5)$   & 1/2   &  0  &  0   & -3/2  \\
\hline
      &    $3(1,1,{\overline 5})$   & -1/2   &  0  &  0   & 3/2  \\
\hline
      &  $3(1,1,10)$    & -1/2   &  0   &  0   &  -1/2  \\
\hline
      &  $(1,1,{\overline {10}})$    & 1/2   &  0   &  0 &  1/2  \\
\hline
   &   $3(1,1,1)$  &  -1/2   & 0   &    0  & -5/2  \\
\hline
   &   $(1,1,1)$  &  1/2   &  0  & 0 & 5/2  \\
\hline
\end{tabular}
\end{center}
\caption{Particle content and charges of the level one
$SU(5)^3$ model.}
\label{tseis}
\end{table}

In this model the $U(1)$ generator $Q_A$ is anomalous whereas the
other three $U(1)$s are anomaly free. Thus, the classical
vacuum is unstable and we have to look for a nearby vacuum which
is one-loop stable. Since $TrQ_A=-48$, some field with
{\it positive $Q_A$ } must acquire a vev to stabilize the
D-term. In general, several fields do acquire vevs to cancel
all the D-terms. We now describe an interesting scenario.
Let us denote by $\eta_1$ the singlet in the
$(\theta,L)$ sector with positive $Q_A$ charge and by
$\eta_2$ the singlet in the $\theta\omega$ sector also with
positive $Q_A$ charge. Then, the following field direction leads
to cancellation of all D-terms and F-terms in the scalar potential
\begin{eqnarray}
& |\eta_1|^2\ =\ {g\over {4\sqrt 2\pi ^2}}M_{string}^2  & \nonumber \\
& |\eta_2|^2\ =\ {g\over {8\sqrt 2\pi ^2}}M_{string}^2  & \nonumber \\
& Tr({\phi }^2-{{\bar {\phi}}^2})\ =\
{{5g}\over {16\sqrt 2\pi ^2}}M_{string}^2 &
\label{svevs}
\end{eqnarray}
where $\phi $ and ${\bar {\phi }}$ denote respectively
the $({\bar 5},5,1)$ and
$(5,{\bar 5},1)$ fields in the $U_3$ untwisted sector.
A diagonal vev $\phi ^i_j=v\delta ^i_j$ would spontaneously
break the first two $SU(5)$ factors down to a diagonal $SU(5)$
model realized at $k=2$. The unbroken gauge group at this level
would be $SU(5)_2\times U(1)_{Q_1+Q_2}\times SU(5)$. Depending
on the value of the field ${\bar {\phi }}$ there may be a direct
breaking from $SU(5)^2$ down to the standard model.

In the process of symmetry breaking some of the untwisted
matter fields get a mass due to the existence of couplings in the
$Z_2\times Z_2$ orbifold of the type $U_1 U_2 U_3$.
The remaining massless states turn out to be
\begin{eqnarray}
& U_1\ : (5,1)_{1,-1}+({\bar 5},5)_{-1,-1}   & \nonumber \\
& U_2\ : ({\bar 5}, 1)_{-1,-1}+(5,{\bar 5})_{1,-1}  &  \nonumber \\
& U_3\ : (24,1)_{0,2}+(1,1)_{0,2}+(1,1)_{0,-2}
 +(1,5)_{0,0}+(1,{\bar 5})_{0,0}    &
\label{restu}
\end{eqnarray}
where the sub-indices are the charges with respect to
$Q_1+Q_2$ and $Q_1-Q_2$ respectively. On the other hand,
the twisted sectors are essentially the same
as in the level one model, with the representations decomposed
in terms of the diagonal $SU(5)$.
Thus we have here a four-generation  $SU(5)$ model
with one adjoint $24$ and several Higgs candidates.
The couplings among these fields are of special interest
for the doublet-triplet problem, as  will be discussed in the next
section. If the vevs of ${\bar {\phi }}$ and ${\phi }$ are of
the same order,
the $SU(5)\times SU(5)$ symmetry could be spontaneously broken
directly to the standard model. The natural scale for this to happen,
as indicated by equation (\ref{svevs}),
would be approximately equal to $ {\sqrt {5g^2/8\pi ^2}} M_{string}$.

There are similar models based on the $Z_2\times Z_4$ orbifold.
For example, the embedding
\begin{eqnarray}
A & = & \frac{1}{2}(1,1,1,-1,-1,0,0,0,0,0,0,0,0,0,0,1) \nonumber \\
B & = &\frac{1}{4}(1,1,1,1,1,1,1,1,1,1,0,0,0,0,0,0)
\end{eqnarray}
leads to the gauge group $SU(5)^2\times SO(10)\times U(1)^3$
and to the untwisted particle content
\begin{eqnarray}
& U_1\ :\ (5,{\bar 5},1)_{1,1,0}+({\bar 5},5,1)_{-1,-1,0}+
(1,1,10)_{0,0,1}+(1,1,10)_{0,0,-1} &  \nonumber \\
& U_2\ :\ (1,5,10)_{0,-1,0}+(5,1,1)_{1,0,1}+(5,1,1)_{1,0,-1} &
\nonumber \\
& U_3\ :\ ({\bar 5},1,10)_{-1,0,0}+(1,{\bar 5},1)_{0,1,1}+
(1,{\bar 5},1)_{0,1,-1}   &
\label{eqn:43}
\end{eqnarray}
where the sub-indices correspond to the charges of the three
$U(1)$s. One linear combination of the three $U(1)$s
is anomalous and a Fayet-Illiopoulos term is generated.
One can check that vevs to $(5,{\bar 5},1)$, $(\bar 5, 5,1)$,
and the positively charged $(1,1,10)$s in the $U_1$ sector,
are D-flat. They trigger the spontaneous breaking of
$SU(5)^2$ down to an $SU(5)$ GUT with one adjoint 24.

In the examples above the GUT Higgs field arises from
the untwisted orbifold sector. Indeed this is the simplest case
but, as emphasized in section 2, fields of the
$(5,{\bar 5})$ form may also appear in a restricted class of twisted
$Z_4$ and $Z_6$ sectors of some orbifolds.
An example is obtained using again the
$Z_2\times Z_4$ orbifold with the same $A$ above but flipping the
sign of the $B$ shift in the entries $6-10$, i.e.,
choosing $B=\frac{1}{4}(1,1,1,1,1,-1,-1,-1,-1,-1,0,0,0,0,0,0)$.
In agreement with our general arguments that
order 4 twists such as $b=1/4(1,-1,0)$ are among the few
that allow massless particles of type $(5,{\bar 5})$,
we find that the $\omega$ sector now includes several copies
of $({\bar 5},5,1)_{{1\over 4},{-{1\over 4}},0}$. Fields
in the opposite representations appear in the $\omega^3$ sector
if one consider a version of the model with a discrete
torsion phase.
Vevs to one of these fields again leads to
an $SU(5)$ at level two and several 24s.
In this case one has $15$s in the untwisted sector. In
fact  by flipping signs in the $B$ as
above we go from a model with an adjoint in the
untwisted sector and $15$s in the twisted sectors to
another model with $15$s in the untwisted sector and
adjoints in the twisted sectors.

There are of course many other models that can be built
but we refrain from presenting further examples.
We have also considered $Z_N\times Z_M$ orbifold
models with discrete torsion. When adding discrete torsion
one goes from one string GUT  to another completely different.
In the $Z_2\times Z_2$ case the addition of discrete
torsion constitutes a sort of mirror operation with
sends families to antifamilies and
viceversa in the $SO(10)$ models constructed. Our main goal
in this paper, however, is
to explain the construction methods and to look for general
patterns that could be common to many string GUT models.
Some of these properties are discussed in the following
section.

\section{Phenomenological aspects of String-GUTs}

In the previous sections we have constructed several different
string-GUTs with gauge groups $SU(5)$ and $SO(10)$ by three
different methods.
Our purpose in this paper is to describe techniques employed
and to try to single out
generic properties of the resulting class of models.
We have shown that the construction of $k=2$ string-GUTs is
relatively easy. In particular, within the context of orbifolds the
complete massless spectrum as well as all the quantum
numbers may be computed without the need of computer help.
Moreover, having all this information
allows us to make a cross-check of the modular invariance of the
theories obtained by verifying the cancellation of
gauge anomalies which may require
the Green-Schwarz mechanism at work.

It is certainly risky to extract general conclusions from a
limited class of models. Some of the general patterns we find are
probably just a consequence of the technique used and it is
non-trivial to separate what is generic in string theory and
what is just a technical property of the method.
Nevertheless we think it is worth to give
a summary of the properties of the
string-GUTs found.

\bigskip

{\bf i.} {\it Generic features of the $SO(10)$ models}
\bigskip

Both $SO(10)$ and $SU(5)$ string GUTs may
be constructed, although $SO(10)$ seems to appear more easily
when  either continuous Wilson lines or permutations are used to
build the specific model. The $SO(10)$ GUTs all share similar
properties: they have a single $54$ Higgs multiplet coming from
the untwisted sector to do the GUT breaking. There are no $45$s. We
believe that this is a rather common feature of {\it left-right symmetric}
$k=2$ string GUTs. This may be partially understood
with the help of Table \ref{tuno}.
Indeed, both representations $45$ and $54$ of $SO(10)$
contribute a large amount to the left-handed conformal
weight $h_{KM}$ in such a way that they are more likely expected
in the untwisted sector of the orbifolds.
Although $45$s of $SO(10)$
may in principle also appear in some $Z_4$ or $Z_6$ twisted sector, we
have not found any example.
In fact, in our approach, GUT Higgses descend from a $(10,10)$
multiplet of an underlying $SO(10)^2$ at level $k=1$
and, since $h_{KM}=1$ for such a multiplet, it cannot be present in
any twisted sector.

In principle, both $45$s and $54$s could
arise  from the untwisted sector. However,
in all the  three different methods discussed
in the text the resulting  $SO(10)$ algebra at $k=2$ is
realized as the diagonal sum of  some $SO(10)^2$,
$k=1$,    subalgebra of
$SO(32)$. This inhibits the presence of a   $45$ in the
untwisted sector. Indeed, if the model is constructed  by a
permutation method, in the decomposition
$(10,10)=45+54+1$ the projection
on permutation-invariant states
will kill the antisymmetric $45$
field. If the method used is any of the other two, one can
understand it, to some extent, as a continuous Higgs mechanism in
which $SO(10)\times SO(10)$ is spontaneously broken to the
diagonal subgroup by a $(10,10)$. Now, the 45 broken generators
are given mass by the antisymmetric piece of the $(10,10)$
and no massless $45$ is then left. Instead, the $54+1$ symmetric
components remain massless. This explains the presence of one massless
$54$ in this type of models.

It is important to remark that the above arguments are no longer
true in the case of left-right asymmetric strings such as
asymmetric orbifolds. In this case, as we mentioned before, there
exists the freedom of twisting the compactified right-movers while
leaving untouched their left-handed counterparts. Therefore,
in $M_L^2$, c.f. eq. (\ref{ml}), there will be no energy shift
$E_0=0$ so that $45$s and $54$s may also surface in any twisted
sector. Indeed, in Ref. \cite{prep} we have constructed explicit
asymmetric orbifolds models in which both $45$s and $54$s
show up in twisted sectors. On the other hand, some of the simple
features of orbifold strings disappear in the asymmetric case,
including the possible interpretation of the 4-D string as
a compactification of a 10-D heterotic string.

Besides the $54$-plet, the $SO(10)$ models do in general contain
$(16+{\overline {16}})$ pairs and both combined can break the symmetry
down to the SM. Candidates for Higgs doublets usually also appear
inside abundant massless
10-plets. Notice that with a 54-plet of Higgs fields the
natural intermediate scale symmetry is the Pati-Salam
$SU(4)\times SU(2)_L\times SU(2)_R$ symmetry.

\bigskip

{\bf ii.} {\it Structure of the GUT-Higgs potential}

\bigskip

In the construction of good-old SUSY-GUTs, the existence of
certain couplings driving a vev for the GUT-Higgs was instrumental.
For example, in SUSY-$SU(5)$ one assumes the existence
of terms in the superpotential
\cite{guts} :
\begin{equation}
W_5\ =\ \Phi _{24}^3\ +\ M\Phi _{24}^2
\label{sup24}
\end{equation}
where $\Phi _{24}$ is the adjoint GUT-Higgs.
This leads to a potential with several degenerate minima
corresponding to $SU(5)$, $SU(4)\times U(1)$ or
$SU(3)\times SU(2)\times U(1)$ symmetries.

In string theory there are no explicit mass
terms: a particle is either massless or has a mass at the string
scale. In this latter case it makes no sense to consider
a particular massive field as
part of the effective Lagrangian while neglecting
many others, thus
the explicit mass term is absent. We also find that in the
class of left-right symmetric string-GUTs we have constructed the
cubic term is also typically absent. In particular, in all
models in which the GUT-Higgs is in the untwisted sector, e.g.
the $54$s in the $SO(10)$ models or the $24$s in the
$SU(5)$ flat-direction models, the GUT-Higgs fields behave
as string moduli and do not have self-interactions at all.
{}From the 4-D point of view this may be understood as a consequence
of the discrete $Z_N$ R-symmetries which originate on the
right-handed part of the string. These also imply the
absence of couplings such as
$((5,{\bar 5})({\bar 5},5))^n$ in $SU(5)^2$ GUTs or
$(10,10)^{2n}$ in
$SO(10)^2$.

If the GUT-Higgs originates in a twisted sector, c.f. the
$Z_2\times Z_4$ example at the end of previous section, the GUT-Higgs
does not need to behave as a string modulus.
Although we have not found any example,
self-interactions of the GUT-Higgs can exist in this case.
However, the presence of GUT-Higgses in twisted
sectors of left-right {\it symmetric} orbifolds is relatively
uncommon, so one may say that in this type of string-GUTs
the absence of self-couplings of the GUT-Higgs is quite generic.
In the case of asymmetric orbifolds GUT-Higgses may appear easily in
twisted sectors and hence they do not necessarily behave like string
moduli.

The absence of explicit mass terms and in some cases even of
cubic terms makes it difficult to obtain GUT-Higgs superpotentials
with the best desirable phenomenological properties. In particular
it will be hard to find string-GUTs in which the particle
content below the unification scale is just that of the
minimal SUSY-SM.

If there are no self-interactions for the GUT-Higgs fields,
there will be some extra matter fields remaining in the
massless spectrum after symmetry breaking. For example,
upon $SU(5)$ symmetry breaking by an adjoint $24$, twelve
out of the 24 fields remain massless. They transform as
\begin{equation}
(8,1,0)\ +\ (1,3,0)\ +\ (1,1,0)
\label{eqn:45}
\end{equation}
under $SU(3)\times SU(2)\times U(1)$. In the case of $SO(10)$, the
extra fields depend on the GUT-Higgs triggering symmetry breaking.
If it is broken by a $54$, the resulting group in a first step
is $SU(4)\times SU(2)_L\times SU(2)_R$ and the
following particles remain light
\begin{equation}
(20,1,1)\ +\ (1,3,3)\ +\ (1,1,1)
\label{eqn:46}
\end{equation}
After further symmetry breaking down to the standard model,
e.g. through a $(16+{\bar {16}})$ pair, those fields transform as
\begin{equation}
(8,1,0)  +  (6,1,-2/3)  +  ({\bar 6},1,2/3)  +
(1,3,0) + (1,3,1) +(1,3,-1)
\label{eqn:47}
\end{equation}
where the third entry gives now the hypercharge.
If $SO(10)$ breaking proceeds through a $45$, the remnant fields would
transform as
\begin{equation}
(8,1,0) + (1,3,0) + (1,1,0) + (1,1,+1) + (1,1,-1)  \ .
\label{eqn:48}
\end{equation}
We see that the different breakings give rise to different extra
matter fields. Since these particles will have masses of the
order of the weak scale, they will sizably contribute to the
running of the gauge coupling constants.
We have performed a one-loop analysis of the running of the
gauge coupling constants and have found that, with the
particle content of the minimal SUSY-SM plus the additional
fields above, there is no
appropriate gauge coupling unification
in the vicinity of $10^{15}-10^{17}$ GeV. Typically
$sin^2\theta _W$ is too large and $\alpha _s$ is too small.
Thus, this class of models cannot break directly to the SM
at a large GUT unification scale. In the case of
$SO(10)$ an intermediate scale of symmetry breaking could
improve the results.

Thus, we see that if there are no self-couplings of the
GUT-Higgs we lose one of the motivations for going to
string-GUTS, a simple (one-step) understanding of
gauge coupling unification.

\bigskip

{\bf iii.} {\it Doublet-triplet splitting and the scalar field moduli
space}

\bigskip
The most severe problem of GUTs is the infamous
doublet-triplet splitting problem of finding a mechanism to
understand why, for example, in the $5$-plet Higgs of $SU(5)$
the Weinberg-Salam doublets remain light while their
coloured triplet partners become heavy enough
to avoid fast proton decay
\cite{guts} . The most simple,
but clearly unacceptable, way to achieve the splitting is
to write a term in the $SU(5)$ superpotential
\begin{equation}
W_H\ =\ \lambda
	H\Phi _{24} {\bar H}\ +\ MH{\bar H}
\label{sup5}
\end{equation}
and fine-tune $\lambda $ and $M$ so that the doublets turn
light and the triplets heavy. Since there are no
explicit mass terms in string theory this inelegant
possibility is not even present. Another alternative
suggested long time ago is
the ``missing partner" mechanism
\cite{miss} . Formulated in
$SU(5)$ it requires the presence of $50$-plets
in the massless sector which is only possible
for level $k \geq 5$, a very unlikely possibility
\cite{fiq,nanop} .

A third mechanism, put forward in the early days of SUSY-GUTs,
is the ``sliding singlet" mechanism
\cite{slid,iruno} . This requires the existence of
a singlet field $X$, with no self-interactions, entering in the
mass term in eq. (\ref{sup5}). $W_H$ is then replaced by
\begin{equation}
W_X\ =\ \lambda
	H\Phi _{24} {\bar H}\ +\ XH{\bar H}   \ .
\label{supx}
\end{equation}
The idea is that the vev of the $24$ is fixed by the potential
in eq. (\ref{sup24}) but the vev
of $X$ is undetermined to start with, i.e. the vev ``slides".
Now, once the electroweak symmetry is broken by the
vevs of $H,{\bar H}$, the minimization conditions give
$\lambda \langle -{3\over 2}v\rangle+\langle X\rangle=0$
where diag($\langle \Phi _{24}\rangle)
=v(1,1,1,-3/2,-3/2)$. In this way
$X$ precisely acquires the vev needed for massless
doublets. This is in principle a nice dynamical
mechanism but it was soon realized that it is
easily spoiled by quantum corrections
\cite{slidprob} . For example, once SUSY is
broken the field $X$ will generically  get a mass $m_X$. If
$m_X^2$ is positive, a large vev
$\langle X\rangle=-\langle\lambda \Phi_{24}\rangle$
will be strongly disfavored. If $m_X^2$ is
negative, a large vev will be preferred,  but in
general not the one that gives massless doublets.  An extra
problem
\cite{nil} for the sliding singlet is that it was
also  shown that light, order weak scale,
singlets  coupling to $H,{\bar H}$ do in
general destabilize the hierarchy by giving a large
$soft$ mass to the Higgs
doublet.

Interestingly enough, we have found that
in string GUTs, couplings of the ``sliding singlet" type are frequent,
the main difference now being that the GUT-Higgs field also ``slides".
In particular, this happens in models in which
the GUT-Higgs is a modulus, as in some of the examples discussed
in the previous sections.
Consider for instance the $SU(5)^3$ model
whose spectrum is shown in Table \ref{tseis}. The untwisted
spectrum of the corresponding level 2 GUT is shown
in eq. (\ref{restu}).
As we mentioned above the couplings of untwisted
fields in the  $Z_2\times Z_2$ orbifold are of
the type $U_1U_2U_3$ and the following terms appear in
the superpotential:
\begin{equation}
 W_X\ =\ (5,1)_{1,-1}\left[ (24,1)_{0,2}\ +\ (1,1)_{0,2}
\right] ({\bar 5},1)_{-1,-1}\ +\ \cdots
\label{eqn:51}
\end{equation}
Vevs of these fields are restricted by absence of an
$SU(5)$ D-term and also we know that
$|(1,1)_{0,2}|^2-|(1,1)_{0,-2}|^2$ is fixed according to
eq. (\ref{svevs}). Otherwise these terms are just like those
in the sliding-singlet mechanism and would in principle give rise
automatically to doublet-triplet spliting, were it not for the
difficulties of the mechanism mentioned above.

Since the scales of the $SU(5)^2\rightarrow SU(5)$, signaled
by $\langle (1,1)_{0,-2}\rangle$,
and $SU(5)\rightarrow SU(3)\times SU(2)\times U(1)$
symmetry-breakings are relatively near, one may alternatively
describe this model as a level one
$SU(5)^2$ model which is broken spontaneously to a
{\it level two} standard model by the two fields
$\phi ^i_j=({\bar 5}, 5)$
and  ${\bar {\phi }}_i^j=( 5,{\bar 5})$ . The sliding-singlet
solution corresponds in this language to the vevs

\begin{equation}
\phi ^i_j =   \left( \begin{array}{ccccc}
		     v & & & & \\
		       &v & & &  \\
		       & & v & & \\
		       & & & v &  \\
		       & & & & v   \end{array}
  \right)\
;\   {\bar {\phi }}_i^j =\left(
		       \begin{array}{ccccc}
		      {\bar v} & & & & \\
		       &{\bar v} & & &  \\
		       & & {\bar v} & & \\
		       & & & 0 &  \\
		       & & & & 0   \end{array}
\right)
\label{pisa}
\end{equation}
Models of this sort have recently been studied by Barbieri et al.
\cite{barb}. The main difference in our case is the absence
both of mass terms and cuartic $\phi ^2{\bar {\phi }}^2$
couplings as  well as the presence of a Fayet-Illiopoulos
term. In the case of Ref. \cite{barb} a vacuum as in
eq. (\ref{pisa}) above may be obtained by fiddling with the
parameters in the potential. In our case the potential is
flat and only non-perturbative  effects could lift the
degeneracy.
{}From the 10 moduli $\phi ^i_i$,  ${\bar {\phi}}_i^i$,
$i=1,..,5$, the two  $\bar \phi_4^4$ and $\bar \phi_5^5$
should remain vev-less for the splitting to occur.
If the sliding-singlet mechanism survived the quantum corrections,
it seems that field configurations with massless Higgs doublets
would be energetically preferred.

Analogous couplings may be found in the $SO(10)$ models with
a massless $54$ behaving as a modulus. Take for example the
first model discussed in section 4 whose massless spectrum is
displayed in table 4. The singlets in the $U_3$ sector $S^0=(1,1)_{0,0}$,
$S^+=(1,1)_{0,1}$, $S^-=(1,1)_{0,-1}$ do also behave as moduli. Both
these singlets and the $54$ couple to the decuplets
$H^+=(10,1)_{0,1}$ and $H^-=(10,1)_{0,-1}$. The sub-indices in all
these fields refer to their $Q$ and $Q_A$ charges. It is easy to check
that there are flat directions in this scalar moduli space
in which the gauge symmetry is broken down to
$SU(4)\times SU(2)\times SU(2)$ and some of the doublets remain light
whereas the colour triplets remain heavy (the symmetry is broken
down to the SM through the vevs of the $16+{\overline {16}}$ pairs). Again,
if the sliding-singlet argument were stable under quantum corrections,
the regions in moduli-space in which there are light doublets
would be energetically favoured.

As the above examples show, the appropriate language to describe
the doublet-triplet splitting problem within the context of the
above string-GUTs is in terms of the scalar moduli space of the
model. At generic points in the moduli space there are no massless
Higgs doublets at all, they are all massive. At some
``multicritical" points in moduli space some Higgs fields
become massless. This is very reminiscent of the behaviour of
the moduli spaces of other well studied string moduli, those
associated to the size and shape of the compact manifold usually
denoted by $T_i$. It is well known that generically there are
points in the $T_i$ moduli space in which extra massless fields
appear. This is also apparently the case of the
moduli space associated to the dilaton complex field $S$. The
problem of understanding the doublet-triplet splitting within this
context would be equivalent to finding out why we are sitting on a
region of moduli space in which massless doublets are obtained.
It could well be that an appropriately modified version of the
sliding-singlet mechanism is at work and that region of moduli
space is energetically favoured.

 \bigskip
{\bf iv.} {\it $SU(5)^n$ and $SO(10)^n$ GUTs}

\bigskip

We have seen that some of the simplest string-GUTs are obtained
by starting with a $SU(5)^2$ or $SO(10)^2$ model at $k=1$ and giving
vevs to $(5,{\bar 5}), ({\bar 5},5)$ or $(10,10)$ fields. It is worth
stressing that this type of structure is very natural from the
point of view of 4-D strings obtained from the $Spin(32)/Z_2$
heterotic theory. Groups with repeated factors. e.g.
$(SU(5)\times U(1))^3\times U(1)_A$, may easily appear as we showed in
section 3.
For particular models some $SU(5)\times U(1)$ factor(s)
may be enhanced to $SO(10)$. The required Higgs fields to break the GUT
symmetry are always present in the untwisted sectors of the above
orbifolds. Replicated GUTs have been recently studied
from a different perspective in ref. \cite{barb}.
Notice however that in our case the couplings for the GUT-Higgs fields
are rather different and, in particular, there are
no self-interactions.

There is a related class of GUT models which also deserves attention.
One may also obtain higher level GUTs by starting with $k=1$ models
with gauge group factors $G_{GUT}\times {\hat G}$ such that
$G_{GUT}\subseteq {\hat G}$. An example of this is the first model
in  section 4. This may be understood as a $k=1$ model with
gauge group $SO(10)\times SO(18)\times U(1)^2$ which is
continuously broken  to $SO(10)_2\times SO(8)\times U(1)^2$ through
appropriate vevs of the multiplet $(10,18)$ in the untwisted
sector.

Models at $k=1$ with gauge group $SU(5)^3$ may be easily constructed.
However it does not seem to be trivial to find flat directions
breaking the symmetry to the diagonal $SU(5)$ subgroup of the three
factors. The reason is that F-term couplings between the untwisted
fields $(5,{\bar 5},1)$, $({\bar 5},1,5)$ and $(1,5,{\bar 5})$
are allowed.

\bigskip

{\bf v.} {\it  The r\^ole of the one-loop Fayet-Illiopoulos term}

\bigskip

As we have seen, a very common feature in string GUTs
is the presence of a one-loop dilaton-dependent
Fayet-Illiopoulos term whenever there is an
extra anomalous $U(1)_A$ in the theory.
As it turns out, most of the models do have
such anomalous $U(1)$s. The existence of the F-I term often has an
important impact in the phenomenology because it forces some
charged massless fields to get vevs thus inducing symmetry
breaking.

We showed some examples in which the F-I term actually
triggers the $G^2\rightarrow G$ symmetry breaking
leading to level two GUTs, and even the breaking down to the
standard model. This is potentially very interesting since
the natural scale of GUT symmetry breaking is then related to the
string scale by one-loop factors, i.e. $M_{GUT}\sim (1/ 8\pi )
M_{string}$. This result is however less appealing if indeed extra
massless  particles remain in the spectrum after symmetry
breaking because then  the computed unification mass will not be the
one of   minimal $SU(5)$, for
instance.

The effects of the F-I term may not always be desired.
Sometimes its presence spoils some otherwise interesting tree level
vacua. For example, we have seen that
models with three GUT factors like $SU(5)^3$ or $SO(10)^3$
can be easily built. However, in some of these models F-I terms
may force the breaking of at least one of the GUT factors.

\section{Final comments and outlook}

In this paper we have tackled the construction of standard
$SO(10)$ and $SU(5)$ GUTs from 4-D string theories. One of our
motivations has been to explore whether the effective low-energy
limit of these 4-D strings resemble the well-known SUSY-GUTs
introduced more than ten years ago or rather, string SUSY-GUTs have
some specific properties on their own. The success of the
SUSY-GUTs prediction for $sin^2\theta _W$ makes this exploration, in
our opinion, worth pursuing. Our study requires the construction
of 4-D string models in which the GUT gauge group is realized
at high ($k>1$)  level, otherwise the Higgs fields
necessary to break the grand unified symmetry cannot be in the string
spectrum.

In our approach, string GUTs are built by employing orbifold
techniques. Within this scheme the massless spectra of the models
may be easily computed without the need of computer help.
The quantum numbers of the massless particles can also be
determined and the cancellation of all anomalies can be verified.
This provides a useful cross-check of the modular
invariance of the models. One can also combine the structure
of the string-GUTs so obtained with several phenomenologically
interesting results, available in the orbifold context, such as
one-loop corrections to coupling constants, SUSY-breaking soft term
computations, etc..

In the present article we have concentrated on the case of
symmetric $(0,2)$ orbifolds and have left the consideration of
more general cases including asymmetric orbifolds for future work.
Thus, the 4-D strings we are constructing may be understood as
compactifications of the 10-dimensional $Spin(32)/Z_2$ and
$E_8\times E_8$ heterotic strings. To derive our models we have had
to extend some of the known results about Abelian orbifolds in
the presence of discrete Wilson lines. In particular, there are
some subtleties concerning the generalized GSO projection
in the presence of Wilson lines which are discussed in the
appendix. We also discuss some aspects of the dependence
of the spectra of $Z_M\times Z_N$ orbifolds on the choice of
underlying compactified lattice.

Three different methods, developed in a previous work, are
used to build our models. The first one
involves turning on continuous Wilson lines when the orbifold
twist is realized in the gauge degrees of freedom by an automorphism
of the gauge lattice. The second method uses the possibility of
embedding discrete (order 2) Wilson lines as a permutation of two
gauge groups of an original $k=1$ model. Finally, a third method
considers flat scalar field directions in which a (semisimple)
$k=1$ gauge group is continuously broken to a subgroup involving
diagonal generators which are realized at $k=2$. There are some
connections among these three methods and sometimes the same
model may be obtained in several possible ways.

In the past, heterotic string compactifications have mostly
been based on the $E_8\times E_8$ theory whereas
the $Spin(32)/Z_2$ theory has been consistently ignored to
the point that it is very difficult to find $(0,2)$ examples
of $Spin(32)/Z_2$-based compactifications in the literature.
Interestingly enough, we find that the $Spin(32)/Z_2$
heterotic theory is the natural starting point in the derivation
of string GUTs. We also find that the replicated GUT groups
$SO(10)^3\times U(1)_A$ and $(SU(5)\times U(1))^3\times U(1)_A$
are naturally embedded into the $SO(32)$ gauge group. Indeed,
some, although not all, of the models we have constructed
may be understood as level one
$SU(5)\times SU(5)$ or $SO(10)\times SO(10)$ GUTs which are
spontaneously broken down to the standard model.

It is possible to make some general model-independent statements
about what string sectors may give rise to the Higgs fields
required for the GUT symmetry breaking. We find that within the
context of symmetric orbifolds those sectors are very much constrained.
For instance, we can show in all generality that the string $SO(10)$
Higgs fields transforming as $45$ or $54$ may only appear either
in the untwisted sector or in a very restricted class of order
4 or 6 twisted sectors. In fact, all the particular $SO(10)$
models constructed have one 54-plet and no adjoint 45-plets.
These constraints are in general relaxed when
asymmetric orbifolds are considered. We remark
that related work using the 4-D strings fermionic formulation
should correspond to asymmetric orbifolds and hence there is no direct
connection between the string-GUTs considered here and those of
Ref. \cite{otros}.

We have attempted to identify some generic features of the string
GUTs obtained through our methods. In many of the examples
the GUT-Higgs fields behave as string moduli, i.e., they have
no self-interactions. This is in general the case when the
GUT-
 reside in the untwisted sector of the orbifold, the most
frequent case in our type of constructions. In any event, it seems
that the absence of GUT-Higgs self-couplings will make rather
difficult to find string-GUTs whose massless sector is just the
MSSM. Typically, extra chiral matter fields, such as color octets and
$SU(2)_L$ triplets, will remain massless. Thus, the presence of
intermediate symmetry-breaking mass scales will be required in order
to be consistent with gauge coupling unification. Other generic
feature which plays a r\^ole in the GUT symmetry breaking is the
presence of one-loop Fayet-Illiopoulos terms.

One of the toughest problems of GUTs in general is the
famous doublet-triplet splitting problem of the Higgs system.
We find that couplings of the ``sliding-singlet" type
are often present in the superpotential. In stringy language,
we find that there are  regions in scalar field moduli space
in which there are light  Higgs doublets and heavy scalar
triplets.  If the sliding-singlet
mechanism were at work, those regions  in moduli space would
be energetically favored. This mechanism was shown
to be generically destroyed by quantum corrections in the
old SUSY-GUT days. It remains to be seen whether strings
provide any improvement over that situation. The fact that
in our class of models the GUT-Higgs often lives in the
untwisted sector of the orbifold which has enhanced $N=4$
supersymmetry could perhaps point in that direction. Recent findings
on the quantum structure of scalar moduli spaces in extended
supersymmetric models could also have an important bearing on this question.

We have only scratched the surface of the class of orbifold string
models leading to SUSY-GUTs at low energies. Many avenues remain
unexplored. We believe that the doublet-triplet splitting problem
is a crucial issue and should be addressed in any model before trying
to extract any further phenomenological consequences such as
fermion masses. It is also important to understand whether it
is possible to build string GUTs in which the massless sector
is just the MSSM, or else whether
the existence of extra massless chiral fields is really generic.
This would dictate the necessity of intermediate scales to
attain coupling constant unification.
All the models displayed have four generations, a result just
due to the particular structure of the $Z_2\times Z_2$ and
$Z_4$ orbifolds which naturally yield even number of generations.
We did not attempt any systematic search for three generation models.
We leave the question
 as well as the construction of models based on asymmetric
orbifolds  to a future
publication \cite{prep}
{}.

\bigskip
\bigskip

\centerline{\bf
Acknowledgements}
 A.F. thanks the Departamento de F\'{\i}sica Te\'orica at UAM  and
the ICTP-Trieste for their hospitality and support while this
work was carried out.
G.A. thanks the Departamento de F\'{\i}sica Te\'orica at UAM for hospitality
and
the Ministry of Education
and Science of Spain (Programa de Cooperaci\'on con Iberoam\'erica)
 and CONICET (Argentina) for financial support. A.M.U.
thanks the Government
of the Basque Country for financial support. This work has been also financed
by
the CICYT (Spain) under grant AEN930673.

\newpage

\appendix
\section{Appendix}

The derivation of the orbifold massless matter content is a
two-step process. First, the states satisfying $M_R=M_L=0$
are found in each sector. Next, the orbifold generalized GSO
projection is imposed. A practical introduction to the full
procedure, for embeddings by shifts, is given in Ref. \cite{fiqs}.
We refer the reader to Appendix A there for the notation
and formulae that we will use in the cases of $Z_2 \times Z_2$
and $Z_4$ relevant to our discussion.

We first consider the case without Wilson lines in which there
are twisted sectors just for each element $g$ belonging to the
Abelian point group ${\cal P}$. The embedding of $g$ is given by
a shift $V_g$. The multiplicity of states
in the $g$-twisted sector is given by an expression of the form
\begin{equation}
D(g) = \frac{1}{|{\cal P}|} \sum_{h \in {\cal P} }  \
\tilde \chi(g,h)  \Delta(g,h)
\label{defd}
\end{equation}
where we have neglected the possibility of discrete torsion for
simplicity. Here $\Delta(g,h)$ are phases that depend on the
sectors and the states. The coefficients $\tilde \chi(g,h)$
are numerical factors that only depend on the sectors. For
instance, $\tilde \chi (1,h) = 1$.

Analysis of the partition function from which $D(g)$ is derived
shows that $\tilde \chi(g,h)$ can generally be written as
\begin{equation}
\tilde \chi(g,h) = {\cal F}(g,h) \ {\cal O}(g,h)
\label{chit}
\end{equation}
Here ${\cal F}(g,h)$ counts the number of inequivalent solutions
for the center of mass $x_{cm}$ that must satisfy the equations
\begin{eqnarray}
x_{cm} & = & gx_{cm} + u \nonumber \\
x_{cm} & = & hx_{cm} + w
\label{xcm}
\end{eqnarray}
where $u,w$ are vectors in the internal six-dimensional lattice
$\Gamma$ with basis $e_i$. For example, as shown
in \cite{nsv}, when $h=1$
\begin{equation}
{\cal F}(g,1) = \left | \frac{N_g}{(1-g)\Gamma} \right |
\label{huno}
\end{equation}
where $N_g$ is the sub-lattice orthogonal to the $g$-invariant
sub-lattice $I_g$. Notice that when $N_g=\Gamma$,
${\cal F}(g,1) = \det (1-g)$ is the number of fixed points of $g$.
When $I_g$ is non-trivial the partition function also includes
an instanton sum. However, we wish to stress that the factor
$1/Vol I_g$ arising from Poisson resummation cancels against
another $Vol I_g$ corresponding to the integral over the
invariant directions of the center of mass.
For $h \not= 1$ we do not have a general formula and a case by
case examination of eq. (\ref{xcm}) is needed.

The other term ${\cal O}(g,h)$ comes from the oscillator piece
in the partition function. Since
there are no fractionally modded oscillators in the non-trivial
directions in $I_g$, a factor has to be extracted out from the
corresponding Theta-functions. More precisely,
\begin{equation}
{\cal O}(g,h) = \frac{1}{\det'_{I_g} (1-h)}
\label{ofac}
\end{equation}
where $\det'_{I_g}$ means that the determinant is evaluated in
the $I_g$ directions with non-zero eigenvalue of $(1-h)$.

Let us now include Wilson lines $L_i$. We recall \cite{imnq}
that in the case of embedding by shifts the $L_i$ are always
quantized. For instance, in a $Z_N$ orbifold it must be that
$NL_i \in \Lambda_{16}$. Also, the $L_i$ are not all independent
but $\forall g \in {\cal P}$ must verify
$gL_i = {\cal M}_{ij} L_j$, where ${\cal M}_{ij}$ is the integer
matrix in $ge_i = {\cal M}_{ij} e_j$.

When the $L_i$ are added, a $g$-twisted sector splits into
several sub-sectors according to how the fixed sets of $g$
detect the Wilson lines. Suppose that a given fixed set,
labelled by $x_g$, is such that $(1-g)x_g = n_i(x_g) e_i$.
The shift associated to $x_g$ is then $[V_g+ n_i(x_g) L_i]$.
Notice that when the extra shift $n_i(x_g) L_i$ happens to
belong to $\Lambda_{16}$ the $x_g$ are Wilson line-blind.
Different $x_g$'s with the same extra shift satisfy the same
mass conditions and can thus be grouped into a sub-sector
in the spectrum.

The Wilson lines also modify the generalized GSO projector
in each sub-sector as we now explain. To each individual
$x_g$ it is convenient to assign a pre-projector
\begin{equation}
D(g|x_g) = \frac{1}{|{\cal P}|} \sum_{h \in {\cal P} }  \
\tilde \chi(g,h|x_g)  \Delta(g,h|x_g)
\label{defdx}
\end{equation}
whose ingredients we now discuss in practical terms. If
$x_g$ is not fixed by $h$, $\tilde \chi(g,h|x_g) $ vanishes
and this $h$ does not contribute to the sum. If $x_g$ is fixed
by $h$, it must be that $(1-h)x_g = m_i(x_g) e_i$. Then,
$\tilde \chi(g,h|x_g) $ is given by a formula such as
(\ref{chit}) with ${\cal F}(g,h)$ essentially being the number
of inequivalent $m_i(x_g)$. In this case, the gauge shifts
$V_g$ and $V_h$ appearing in $ \Delta(g,h|x_g)$ must include
extra contributions given respectively by
$S_g(x_g) = n_i(x_g) L_i$ and $S_h(x_g) = m_i(x_g) L_i$.
We will then use the more explicit notation
\begin{equation}
\Delta(g,h|x_g) \equiv \Delta(g,h|V_g +S_g(x_g) , V_h + S_h(x_g))
= \Delta(g,h) \left |
\begin{array}{l}
 \\  \\
V_g \to V_g + S_g(x_g) \\
V_h \to V_h + S_h(x_g) \
\end{array}  \right.
\end{equation}
Modular invariance constraints on the $L_i$ result from their
presence in the phases $\Delta$. For instance, a full shift
$(V_g + S_g)$ must satisfy a condition similar to (\ref{Vcond}).

Once the quantization and relations among the $L_i$ are taken
into account, it is generally found that several $x_g$ have
the same $S_g$ up to $\Lambda_{16}$ lattice vectors.
We then define a sub-sector $(g,S_g)$ with overall projector
\begin{equation}
D(g,S_g) =
\sum_{x_g \, |  \, n_i(x_g) L_i \, \equiv \, S_g} \ D(g|x_g)
\label{defdgs}
\end{equation}
These issues will be clarified in specific examples below.

\subsection{$Z_2 \times Z_2$}

We take $\Gamma$ to be the hypercubic $SO(4)^3$ lattice. The
elements of the point group are $\{ 1, \theta, \omega,
\theta\omega \}$. The corresponding shifts are
\begin{eqnarray}
\theta :
& a=(\frac{1}{2},0,-\frac{1}{2}) \longrightarrow A & \nonumber \\
\omega :
& b=(0,\frac{1}{2},-\frac{1}{2}) \longrightarrow B & \nonumber \\
\theta\omega :
& c=(\frac{1}{2},-\frac{1}{2},0) \longrightarrow C &
\label{abc}
\end{eqnarray}
where $2A \, , \, 2B \, \in \Lambda_{16}$ and $C=A-B$.

There are several allowed forms for the twists $\theta$ and
$\omega$ written in a six dimensional basis $(1,0,0,0,0,0)$,
etc. \cite{orb2}. To write these twists
we use block notation, also $\sigma_1$ is the $2\times 2$
Pauli matrix. One possibility is
\begin{eqnarray}
&\theta=diag \ (-1,\sigma_1, \sigma_1)  \  ;  \
\omega=diag \ (-\sigma_1,-1, -\sigma_1) & \nonumber \\
&\tilde \chi(\theta,1) = \tilde \chi(\theta,\theta) =4 \  ;  \
\tilde \chi(\theta,\omega) = \tilde \chi(\theta,\theta\omega) = 2 &
\label{puno}
\end{eqnarray}
Among others, another possibility is
\begin{eqnarray}
&\theta=diag \ (-1,1,-1)   \  ;  \
\omega=diag \ (1,-1,-1) & \nonumber \\
&\tilde \chi(\theta,1) = \tilde \chi(\theta,\theta) =
\tilde \chi(\theta,\omega) = \tilde \chi(\theta,\theta\omega) = 16 &
\label{pdos}
\end{eqnarray}
In the following we will take choice (\ref{puno})
since it leads to lower multiplicities.

In this orbifold the untwisted sector splits in three
sub-sectors denoted $U_i$. The corresponding projections are
\begin{eqnarray}
U_1 : & P\cdot A = \frac{1}{2} + \ int\ , \ P\cdot B = int & \nonumber \\
U_2 : & P\cdot A = int\ , \ P\cdot B = \frac{1}{2} + \ int  &
\nonumber \\
U_3 : & P\cdot A = \frac{1}{2} + \ int\ , \ P\cdot B = \frac{1}{2} + \ int &
\label{usec}
\end{eqnarray}
Each allowed state appears in one copy.

The multiplicity of states in each twisted sector is obtained by
substituting (\ref{puno}) into (\ref{defd}). The results are
\begin{eqnarray}
D(\theta)& = &\frac{1}{2} \left [ 2 + 2 \Delta(\theta,\theta) +
 \Delta(\theta,\omega) + \Delta(\theta,\theta\omega) \right ]
\nonumber \\
D(\omega)& = &\frac{1}{2} \left [ 2 +  \Delta(\omega,\theta) +
2 \Delta(\omega,\omega) + \Delta(\omega,\theta\omega) \right ]
\nonumber \\
D(\theta\omega)& = &\frac{1}{2} \left [ 2 +
\Delta(\theta\omega,\theta) + \Delta(\theta\omega,\omega) +
2 \Delta(\theta\omega,\theta\omega) \right ]
\label{abcpro}
\end{eqnarray}
where $\Delta(\theta^k\omega^l,\theta^t\omega^s) =
\Delta(k,l;t,s)$ in the notation of \cite{fiqs}.

Now consider turning on the Wilson lines $L_1=L_2 =L$ with
$2L \in \Lambda_{16}$. The $\theta$ sector splits into
sub-sectors with embeddings $A$ and $A+L$. In the $(\theta,0)$
sector the fixed sets are
$x_1 = (0,0)\otimes (\alpha,\alpha) \otimes (\beta, \beta)$ and
$x_2 = (\frac{1}{2},\frac{1}{2})\otimes (\alpha,\alpha)
\otimes (\beta, \beta)$. They are fixed by $\omega$ and
$\theta\omega$ for the particular values $\alpha=0, \frac{1}{2}$
and $\beta=0,\frac{1}{2}$. The overall projector then turns out
to be
\begin{equation}
D(\theta,0) = \frac{1}{2} \left [ 1 +  \Delta(\theta,\theta|A,A) +
 \Delta(\theta,\omega|A,B) + \Delta(\theta,\theta\omega|A,C) \right ]
\label{asec}
\end{equation}
In the $(\theta,L)$ sector the fixed sets are
$x_3 = (\frac{1}{2},0)\otimes (\alpha,\alpha) \otimes (\beta, \beta)$ and
$x_4 = (0,\frac{1}{2})\otimes (\alpha,\alpha)
\otimes (\beta, \beta)$. They are not fixed by either $\omega$
or $\theta\omega$. Hence,
\begin{equation}
D(\theta,L) = \frac{1}{2} \left [ 1 +
\Delta(\theta,\theta | A+L,A+L) \right ]
\label{alsec}
\end{equation}

Finally, the sectors $\omega$ and $\theta\omega$ do not split at all
and furthermore, the projectors remain those given in
eq. (\ref{abcpro}).

\subsection{$Z_4$}

We take $\Gamma$ to be the product of two $SU(4)$ root lattices.
The point group is generated by the order four element $\theta$
realized with shifts
\begin{equation}
\theta : v=(\frac{1}{4}, \frac{1}{4}, - \frac{1}{2}) \longrightarrow V
\end{equation}
In each $SU(4)$, $\theta$ is represented by the Coxeter rotation
given by
\begin{equation}
\theta_c = \left (
\begin{array}{ccc}
0 & 0 & -1 \\
1 & 0 & -1 \\
0 & 1 & -1 \end{array}
\right )
\label{cox}
\end{equation}
when written in the root basis $\{e_1, e_2, e_3\}$. It is easy to
see that $\theta_c$ fixes the origin $w_0=0$ plus the three
fundamental weights $w_1, w_2, w_3$. Hence, $\theta$ has altogether
16 fixed points that are also fixed by $\theta^2$. We then find
\begin{equation}
\tilde \chi(\theta,1) = \tilde \chi(\theta,\theta) =
\tilde \chi(\theta,\theta^2) = \tilde \chi(\theta,\theta^3) = 16
\label{vsec}
\end{equation}
The sector $\theta^2$ is subtler since there are fixed tori.
Indeed, $\theta_c$ has two fixed directions
$u_0 = \alpha(e_1 + e_3)$ and $u_1=w_1 + \alpha(e_1 + e_3)$.
Both are fixed by $\theta_c$ provided $\alpha=0, \frac{1}{2}$.
This is another way of saying that $\theta_c$ and $\theta_c^2$
simultaneously leave fixed the four points $w_i$. Then,
\begin{equation}
\tilde \chi(\theta^2,1) = \tilde \chi(\theta^2,\theta^2) = 4 \  ;  \
\tilde \chi(\theta^2,\theta) = \tilde \chi(\theta^2,\theta^3) =
\frac{16}{4} = 4
\label{2vsec}
\end{equation}
where we have used that
$(e_1+e_3)$ is eigenvector of $(1-\theta_c)$ with eigenvalue 2.

The projections in the untwisted sector are
\begin{eqnarray}
U_1, U_2 : & P\cdot V = \frac{1}{4} + \ int \nonumber \\
U_3 : & P\cdot V = \frac{1}{2} + \ int
\label{u4sec}
\end{eqnarray}
Each allowed state has multiplicity one. The multiplicity of the
twisted sectors is obtained by substituting (\ref{vsec}) and
(\ref{2vsec}) in (\ref{defd}). For example,
\begin{equation}
D(\theta^2) = 1 + \Delta(\theta^2,\theta) +
\Delta(\theta^2,\theta^2) + \Delta(\theta^2,\theta^3)
\end{equation}
This result was also obtained in Ref. \cite{ek} by slightly
different arguments.

Let us now turn on Wilson lines. These can be of order two or order
four. We will analyze the latter case. We then consider
$L_1 = L_2 = L_3 = L$ with $4L \in \Lambda_{16}$. The $\theta$
sector splits into sub-sectors with shifts $V, V \pm L$ and
$V+2L$. The $\theta^2$ sector splits into sub-sectors with
shifts $2V$ and $2V+2L$.

The Wilson line also affects the generalized GSO projection,
even when it does not appear in the shift.
In order to make this issue clearer, let us show this projector
explicitly  in the $(\theta^2,0)$ sector.
The fixed sets with no extra shift are $u_0 \otimes u_i$
which, as mentioned before, are fixed by $\theta$ for
particular values of $\alpha$. For $\alpha=0$,
$S_{\theta}(u_0\otimes u_i) = S_{\theta}(w_0\otimes u_i) = 0$.
For $\alpha=\frac{1}{2}$,
$S_{\theta}(u_0\otimes u_i) = S_{\theta}(w_2\otimes u_i) = 2L$.
We then find
\begin{eqnarray}
D(\theta^2,0) & & =  \frac{1}{4} \big \{ 2 +
\big [  \Delta(\theta^2, \theta | 2V,V) +
\Delta(\theta^2, \theta | 2V,V+2L) \big ]  +
2 \, \Delta(\theta^2, \theta^2 | 2V,2V) +  \nonumber \\
  &  &
\big [  \Delta(\theta^2, \theta^3 | 2V,3V) +
\Delta(\theta^2, \theta^3 | 2V,3V+2L) \big ]  \big \}
\label{d2v}
\end{eqnarray}
The effect of $L$ filtrates through phases $\Delta$.

\end{document}